\PassOptionsToPackage{table,xcdraw}{xcolor}
\documentclass[aps,prd,twocolumn,superscriptaddress,nofootinbib,orcidlink]{revtex4-2}
\usepackage{orcidlink}
\usepackage{amsmath}
\usepackage{graphicx}
\usepackage{array}
\usepackage{xcolor}
\usepackage{amssymb}
\usepackage{float}
\usepackage{placeins}
\usepackage{multirow}
\usepackage{booktabs}
\usepackage{microtype}
\usepackage{textcomp} %
\usepackage[justification=justified,singlelinecheck=false]{caption}

\usepackage{ragged2e} %
\makeatletter
\long\def\@makecaption#1#2{%
  \vskip\abovecaptionskip
  {\small \justifying #1: #2\par}%
  \vskip\belowcaptionskip}
\makeatother

\begin{document}

\title{Search for the $^{16}\text{O}(ppp) \rightarrow ^{13}\text{C} \pi^+ \pi^+ e^+$ Decay Mode in Super-Kamiokande Using Machine Learning Techniques}

\def\SubmissionDate{20260922}
\newcommand{\AFFicrr}{\affiliation{Kamioka Observatory, Institute for Cosmic Ray Research, University of Tokyo, Kamioka, Gifu 506-1205, Japan}}
\newcommand{\AFFkashiwa}{\affiliation{Research Center for Cosmic Neutrinos, Institute for Cosmic Ray Research, University of Tokyo, Kashiwa, Chiba 277-8582, Japan}}
\newcommand{\AFFipmu}{\affiliation{Kavli Institute for the Physics and
Mathematics of the Universe (WPI), The University of Tokyo Institutes for Advanced Study,
University of Tokyo, Kashiwa, Chiba 277-8583, Japan }}
\newcommand{\AFFmad}{\affiliation{Department of Theoretical Physics, University Autonoma Madrid, 28049 Madrid, Spain}}
\newcommand{\AFFubc}{\affiliation{Department of Physics and Astronomy, University of British Columbia, Vancouver, BC, V6T1Z4, Canada}}
\newcommand{\AFFbu}{\affiliation{Department of Physics, Boston University, Boston, MA 02215, USA}}
\newcommand{\AFFuci}{\affiliation{Department of Physics and Astronomy, University of California, Irvine, Irvine, CA 92697-4575, USA }}
\newcommand{\AFFcsu}{\affiliation{Department of Physics, California State University, Dominguez Hills, Carson, CA 90747, USA}}
\newcommand{\AFFcnm}{\affiliation{Institute for Universe and Elementary Particles, Chonnam National University, Gwangju 61186, Korea}}
\newcommand{\AFFduke}{\affiliation{Department of Physics, Duke University, Durham NC 27708, USA}}
\newcommand{\AFFgifu}{\affiliation{Department of Physics, Gifu University, Gifu, Gifu 501-1193, Japan}}
\newcommand{\AFFgist}{\affiliation{GIST College, Gwangju Institute of Science and Technology, Gwangju 500-712, Korea}}
\newcommand{\AFFuh}{\affiliation{Department of Physics and Astronomy, University of Hawaii, Honolulu, HI 96822, USA}}
\newcommand{\AFFicl}{\affiliation{Department of Physics, Imperial College London , London, SW7 2AZ, United Kingdom }}
\newcommand{\AFFkek}{\affiliation{High Energy Accelerator Research Organization (KEK), Tsukuba, Ibaraki 305-0801, Japan }}
\newcommand{\AFFkobe}{\affiliation{Department of Physics, Kobe University, Kobe, Hyogo 657-8501, Japan}}
\newcommand{\AFFkyoto}{\affiliation{Department of Physics, Kyoto University, Kyoto, Kyoto 606-8502, Japan}}
\newcommand{\AFFliv}{\affiliation{Department of Physics, University of Liverpool, Liverpool, L69 7ZE, United Kingdom}}
\newcommand{\AFFmiyagi}{\affiliation{Department of Physics, Miyagi University of Education, Sendai, Miyagi 980-0845, Japan}}
\newcommand{\AFFnagoya}{\affiliation{Institute for Space-Earth Environmental Research, Nagoya University, Nagoya, Aichi 464-8602, Japan}}
\newcommand{\AFFkmi}{\affiliation{Kobayashi-Maskawa Institute for the Origin of Particles and the Universe, Nagoya University, Nagoya, Aichi 464-8602, Japan}}
\newcommand{\AFFpol}{\affiliation{National Centre For Nuclear Research, 02-093 Warsaw, Poland}}
\newcommand{\AFFsuny}{\affiliation{Department of Physics and Astronomy, State University of New York at Stony Brook, NY 11794-3800, USA}}
\newcommand{\AFFokayama}{\affiliation{Department of Physics, Okayama University, Okayama, Okayama 700-8530, Japan }}
\newcommand{\AFFosaka}{\affiliation{Department of Physics, Osaka University, Toyonaka, Osaka 560-0043, Japan}}
\newcommand{\AFFox}{\affiliation{Department of Physics, Oxford University, Oxford, OX1 3PU, United Kingdom}}
\newcommand{\AFFqmul}{\affiliation{School of Physics and Astronomy, Queen Mary University of London, London, E1 4NS, United Kingdom}}
\newcommand{\AFFregina}{\affiliation{Department of Physics, University of Regina, 3737 Wascana Parkway, Regina, SK, S4SOA2, Canada}}
\newcommand{\AFFseoul}{\affiliation{Department of Physics and Astronomy, Seoul National University, Seoul 151-742, Korea}}
\newcommand{\AFFsheff}{\affiliation{School of Mathematical and Physical Sciences, University of Sheffield, S3 7RH, Sheffield, United Kingdom}}
\newcommand{\AFFshizuokasc}{\affiliation{Department of Informatics in
Social Welfare, Shizuoka University of Welfare, Yaizu, Shizuoka, 425-8611, Japan}}
\newcommand{\AFFstfc}{\affiliation{STFC, Rutherford Appleton Laboratory, Harwell Oxford, and Daresbury Laboratory, Warrington, OX11 0QX, United Kingdom}}
\newcommand{\AFFskk}{\affiliation{Department of Physics, Sungkyunkwan University, Suwon 440-746, Korea}}
\newcommand{\AFFtodai}{\affiliation{Department of Physics, University of Tokyo, Bunkyo, Tokyo 113-0033, Japan }}
\newcommand{\AFFtit}{\affiliation{Department of Physics, Institute of Science Tokyo, Meguro, Tokyo 152-8551, Japan }}
\newcommand{\AFFtus}{\affiliation{Department of Physics and Astronomy, Faculty of Science and Technology, Tokyo University of Science, Noda, Chiba 278-8510, Japan }}
\newcommand{\AFFtriumf}{\affiliation{TRIUMF, 4004 Wesbrook Mall, Vancouver, BC, V6T2A3, Canada }}
\newcommand{\AFFtokai}{\affiliation{Department of Physics, Tokai University, Hiratsuka, Kanagawa 259-1292, Japan}}
\newcommand{\AFFtsinghua}{\affiliation{Department of Engineering Physics, Tsinghua University, Beijing, 100084, China}}
\newcommand{\AFFynu}{\affiliation{Department of Physics, Yokohama National University, Yokohama, Kanagawa, 240-8501, Japan}}
\newcommand{\AFFllr}{\affiliation{Ecole Polytechnique, IN2P3-CNRS, Laboratoire Leprince-Ringuet, F-91120 Palaiseau, France }}
\newcommand{\AFFbari}{\affiliation{ Dipartimento Interuniversitario di Fisica, INFN Sezione di Bari and Universit\`a e Politecnico di Bari, I-70125, Bari, Italy}}
\newcommand{\AFFnapoli}{\affiliation{Dipartimento di Fisica, INFN Sezione di Napoli and Universit\`a di Napoli, I-80126, Napoli, Italy}}
\newcommand{\AFFroma}{\affiliation{INFN Sezione di Roma and Universit\`a di Roma ``La Sapienza'', I-00185, Roma, Italy}}
\newcommand{\AFFpadova}{\affiliation{Dipartimento di Fisica, INFN Sezione di Padova and Universit\`a di Padova, I-35131, Padova, Italy}}
\newcommand{\AFFkeio}{\affiliation{Department of Physics, Keio University, Yokohama, Kanagawa, 223-8522, Japan}}
\newcommand{\AFFwinnipeg}{\affiliation{Department of Physics, University of Winnipeg, MB R3J 3L8, Canada }}
\newcommand{\AFFkcl}{\affiliation{Department of Physics, King's College London, London, WC2R 2LS, UK }}
\newcommand{\AFFwarwick}{\affiliation{Department of Physics, University of Warwick, Coventry, CV4 7AL, UK }}
\newcommand{\AFFral}{\affiliation{Rutherford Appleton Laboratory, Harwell, Oxford, OX11 0QX, UK }}
\newcommand{\AFFwu}{\affiliation{Faculty of Physics, University of Warsaw, Warsaw, 02-093, Poland }}
\newcommand{\AFFbcit}{\affiliation{Department of Physics, British Columbia Institute of Technology, Burnaby, BC, V5G 3H2, Canada }}
\newcommand{\AFFtohoku}{\affiliation{Department of Physics, Faculty of Science, Tohoku University, Sendai, Miyagi, 980-8578, Japan }}
\newcommand{\AFFicise}{\affiliation{Institute For Interdisciplinary Research in Science and Education, ICISE, Quy Nhon, 55121, Vietnam }}
\newcommand{\AFFilance}{\affiliation{ILANCE, CNRS - University of Tokyo International Research Laboratory, Kashiwa, Chiba 277-8582, Japan}}
\newcommand{\AFFibs}{\affiliation{Center for Underground Physics, Institute for Basic Science (IBS), Daejeon, 34126, Korea}}
\newcommand{\AFFglasgow}{\affiliation{School of Physics and Astronomy, University of Glasgow, Glasgow, Scotland, G12 8QQ, United Kingdom}}
\newcommand{\AFFoecu}{\affiliation{Media Communication Center, Osaka Electro-Communication University, Neyagawa, Osaka, 572-8530, Japan}}
\newcommand{\AFFminn}{\affiliation{School of Physics and Astronomy, University of Minnesota, Minneapolis, MN  55455, USA}}
\newcommand{\AFFsilesia}{\affiliation{August Che\l{}kowski Institute of Physics, University of Silesia in Katowice, 75 Pu\l{}ku Piechoty 1, 41-500 Chorz\'{o}w, Poland}}
\newcommand{\AFFtoyama}{\affiliation{Faculty of Science, University of Toyama, Toyama City, Toyama 930-8555, Japan}}
\newcommand{\AFFbmcc}{\affiliation{Science Department, Borough of Manhattan Community College / City University of New York, New York, New York, 1007, USA.}}
\newcommand{\AFFnumazu}{\affiliation{National Institute of Technology, Numazu College, Numazu, Shizuoka 410-8501, Japan}}
\newcommand{\AFFniihama}{\affiliation{National Institute of Technology, Niihama College, Niihama, Ehime  792-8580, Japan}}
\newcommand{\AFFucas}{\affiliation{School of Physical Sciences, University of Chinese Academy of Sciences, Beijing 101408, China}}

\AFFicrr
\AFFkashiwa
\AFFmad
\AFFbmcc
\AFFbu
\AFFbcit
\AFFuci
\AFFcsu
\AFFucas
\AFFcnm
\AFFduke
\AFFllr
\AFFgifu
\AFFgist
\AFFglasgow
\AFFuh
\AFFibs
\AFFicise
\AFFicl
\AFFbari
\AFFnapoli
\AFFpadova
\AFFroma
\AFFilance
\AFFkeio
\AFFkek
\AFFkcl
\AFFkobe
\AFFkyoto
\AFFliv
\AFFminn
\AFFmiyagi
\AFFnagoya
\AFFkmi
\AFFpol
\AFFniihama
\AFFnumazu
\AFFsuny
\AFFokayama
\AFFoecu
\AFFox
\AFFral
\AFFseoul
\AFFsheff
\AFFshizuokasc
\AFFsilesia
\AFFstfc
\AFFskk
\AFFtohoku
\AFFtodai
\AFFipmu
\AFFtit
\AFFtus
\AFFtoyama
\AFFtriumf
\AFFtsinghua
\AFFwu
\AFFwarwick
\AFFwinnipeg
\AFFynu

\author{J.~Feng}
\AFFkyoto

\author{K.~Abe \orcidlink{0009-0000-9620-788X}}
\AFFicrr
\AFFipmu
\author{Y.~Asaoka \orcidlink{0000-0001-6440-933X}}
\AFFicrr
\AFFipmu
\author{M.~Harada \orcidlink{0000-0003-3273-946X}}
\AFFicrr
\author{Y.~Hayato \orcidlink{0000-0002-8683-5038}}
\AFFicrr
\AFFipmu
\author{K.~Hiraide \orcidlink{0000-0003-1229-9452}}
\AFFicrr
\AFFipmu
\author{T.~H.~Hung}
\AFFicrr
\author{K.~Ieki \orcidlink{0000-0002-7791-5044}}
\author{M.~Ikeda \orcidlink{0000-0002-4177-5828}}
\AFFicrr
\AFFipmu
\author{J.~Kameda}
\AFFicrr
\AFFipmu
% ...
\ifnum\SubmissionDate>20260921
\author{Z.~Utagawa \orcidlink{0009-0007-3304-9977}}
\AFFkeio
\fi
\author{M.~Fukazawa}
\author{S.~Horiuchi \orcidlink{0009-0005-9007-0700}}
\author{A.~Kawabata \orcidlink{0009-0002-5162-3892}}
\author{Y.~M.~Liu}
\author{Y.~Maekawa \orcidlink{0000-0001-9783-7656}}
\author{Y.~Nishimura \orcidlink{0000-0002-7666-3789}}
\author{A.~Oka}
\AFFkeio

\author{R.~Akutsu}
\author{M.~Friend}
\author{T.~Hasegawa \orcidlink{0000-0002-2967-1954}} 
\author{Y.~Hino \orcidlink{0000-0002-7480-463X}}
\author{T.~Ishida}
\author{T.~Kobayashi} 
\author{T.~Matsubara \orcidlink{0000-0003-3187-6710}}
\author{T.~Nakadaira} 
\AFFkek 
\author{Y.~Oyama \orcidlink{0000-0002-1689-0285}} 
\author{A.~Portocarrero Yrey}
\author{K.~Sakashita} 
\author{T.~Sekiguchi} 
\AFFkek 

\author{N.~Bhuiyan \orcidlink{0009-0002-1227-1548}}
\author{F.~Di Lodovico \orcidlink{0000-0003-3952-2175}}
\author{T.~Katori \orcidlink{0000-0002-9429-9482}}
\author{R.~Kralik \orcidlink{0000-0001-7557-5085}}
\author{N.~Latham \orcidlink{0000-0003-1329-8013}}
\author{R.~M.~Ramsden \orcidlink{0009-0005-3298-6593}}
\author{V.~Siccardi}
\AFFkcl

\author{S.~Aoyama}
\author{H.~Bambara}
\author{Y.~Inaba}
\author{H.~Ito \orcidlink{0000-0003-1029-5730}}
\author{M.~Nishigami}
\author{T.~Sone}
\author{A.~T.~Suzuki}
\AFFkobe
\author{Y.~Takeuchi \orcidlink{0000-0002-4665-2210}}
\AFFkobe
\AFFipmu
\author{S.~Wada}
\author{H.~Zhong}
\AFFkobe

\author{L.~Feng}
\author{S.~Han \orcidlink{0009-0002-8908-6922}}
\author{J.~Hikida}
\author{N.~Fujimoto}
\AFFkyoto
\author{M.~Kawaue \orcidlink{0000-0002-7049-6668}}
\author{T.~Kikawa}
\author{F.~Nakanishi \orcidlink{0000-0003-4408-6929}}
\AFFkyoto
\author{T.~Nakaya \orcidlink{0000-0003-3040-4674}}
\AFFkyoto
\AFFipmu
\author{R.~A.~Wendell \orcidlink{0000-0002-0969-4681}}
\AFFkyoto
\AFFipmu

\ifnum\SubmissionDate>20260914
\author{U.~Limbu}
\AFFliv
\fi
\author{S.~J.~Jenkins \orcidlink{0000-0002-0982-8141}}
\author{N.~McCauley \orcidlink{0000-0002-5982-5125}}
\AFFliv

\author{M.~Fan\`{i} \orcidlink{0000-0002-4284-9614}}
\author{M.~J.~Wilking \orcidlink{0000-0002-9441-7274}}
\author{Z.~Xie \orcidlink{0009-0003-0144-2871}}
\AFFminn

\author{Y.~Fukuda \orcidlink{0000-0003-2660-1958}}
\AFFmiyagi

\author{H.~Menjo \orcidlink{0000-0001-8466-1938}}
\AFFnagoya
\AFFkmi
\author{Y.~Yoshioka}
\AFFnagoya

\author{J.~Lagoda}
\author{J.~Zalipska}
\AFFpol

% ...

\collaboration{The Super-Kamiokande Collaboration}
\noaffiliation

\date{September 22, 2026}

\begin{abstract}
We report a new partial lifetime limit of $4.2 \times 10^{32}$ years for the trinucleon decay mode $^{16}\text{O}(ppp) \rightarrow ^{13}\text{C} \pi^+ \pi^+ e^+$, obtained from a search conducted using the Super-Kamiokande detector with 0.401 megaton-years of exposure across five operational periods (SK-I: 1996--2001, SK-II: 2002--2005, SK-III: 2006--2008, SK-IV: 2008--2018, SK-V: 2019--2020). This represents an improvement of six orders of magnitude over previous experimental constraints. The analysis utilizes a convolutional neural network (CNN) incorporating an attention mechanism—a computational technique that enables the model to focus on the most relevant regions of Cherenkov ring patterns—to enhance event classification, thereby improving the sensitivity of the search. This is the first application of a CNN to a nucleon decay search in Super-Kamiokande. Furthermore, the large dataset available in Super-Kamiokande (hereafter "SK") strengthens the statistical power of the study, enabling a more stringent constraint than those set by prior experiments.
\end{abstract}

\maketitle

\section{Introduction}

The conservation of baryon number, $B$, is a cornerstone of the Standard Model (SM) of particle physics, though it is not dictated by a fundamental gauge symmetry. Its violation is of theoretical interest, particularly in the context of explaining the observed matter-antimatter asymmetry in the universe, as proposed by Sakharov~\cite{sakharov1967}. Grand Unified Theories (GUTs) and other extensions of the SM naturally predict baryon number violation~\cite{georgi1974}, leading to processes such as proton, dinucleon, and trinucleon decays. Because the SM contains no perturbative interaction that mediates proton decay, observation of such a process would constitute evidence for physics beyond the SM. Experimental searches have so far constrained the proton lifetime to be greater than \( 10^{34} \) years~\cite{takenaka2020}, demonstrating the extreme rarity of these decays and the challenges associated with their detection.

Certain extensions of the SM possess an anomaly-free $Z_6$ symmetry, itself a subgroup of the $U(1)_{2Y-B+3L}$ gauge group. 
This symmetry imposes a constraint on allowed processes~\cite{babu2003}:

\[
2\Delta Y - \Delta B + 3\Delta L = 0 \ (\text{mod}\ 6).
\]

\noindent Here, $Y$ denotes the hypercharge, $\Delta Y = 0$, and $L$ denotes the lepton number. The modulo 6 condition arises from the cyclic nature of the $Z_6$ symmetry group, which has order 6, requiring all physical transformations to be invariant under six-fold iterations. This mathematical constraint allows for various values of baryon number violation, such as $\Delta B = 3$, 9, or other values that satisfy the modulo 6 condition. In this study, we focus specifically on processes with $\Delta B = 3$, which can occur through dimension-15 operators and represent the lowest-order baryon number violation permitted by this symmetry.

This study focuses on a specific $\Delta B = 3$ process: trinucleon decay in oxygen nuclei. In this process, $^{16}\text{O}(ppp)$ undergoes decay into $^{13}\text{C} \pi^+ \pi^+ e^+$. Three protons simultaneously decay, producing two $\pi^+$ mesons and one positron, leaving a $^{13}\text{C}$ nucleus as the residual. Throughout this paper, this process is abbreviated as $ppp \rightarrow \pi^+ \pi^+ e^+$. Theoretical predictions for such processes estimate their partial lifetime to be on the order of $10^{33}$ years~\cite{babu2003}. Within the anomaly-free $Z_6$ baryon parity framework described in Ref.~\cite{babu2003}, these $\Delta B = 3$ operators arise as the lowest-dimensional baryon-number-violating interactions that remain consistent with an underlying $U(1)_{2Y-B+3L}$ gauge symmetry, while automatically forbidding dangerous $\Delta B = 1$ and $\Delta B = 2$ transitions even when the new-physics scale lies near the TeV range. The resulting dimension-15 operators respect $B-L$ conservation and can be embedded in unified constructions that simultaneously address neutrino masses and the origin of the baryon asymmetry, thereby linking trinucleon decay searches to the broader program of testing baryogenesis mechanisms resilient to electroweak sphaleron washout. These theoretical motivations make the $ppp \rightarrow \pi^+ \pi^+ e^+$ channel an especially sensitive probe of baryon-parity-protected scenarios.

Previous experiments have investigated baryon-number-violating processes, including searches for trinucleon decay. For instance, GERDA ($\tau > 4.7 \times 10^{25}$ years)~\cite{gerda2023} and the Majorana Demonstrator ($\tau > 1.6 \times 10^{26}$ years)~\cite{alvis2019} have placed constraints on the specific trinucleon decay mode \( ppp \to \pi^+ \pi^+ e^+ \). However, these experiments employed different detection techniques and target materials compared to water Cherenkov detectors. The SK detector, with its large fiducial volume and long exposure time, enables further investigation of this decay mode under different experimental conditions~\cite{fukuda2003}. 

Hadronic final state interactions (hereafter "FSI") within the oxygen nucleus can alter the observed event topology from the initial particle kinematics of both signal and background events, affecting event reconstruction and the performance of conventional selection methods. To address this, the analysis applies a CNN architecture with an attention mechanism to improve event classification. This approach enhances the identification of event signatures affected by FSI and improves background suppression.

This paper is organized as follows. Section~\ref{sec:detector} describes the SK detector and data acquisition. Section~\ref{sec:simulation} presents the signal and atmospheric-neutrino simulations. Section~\ref{sec:analysis} describes event reconstruction, pre-selection, and the CNN analysis. Section~\ref{sec:systematics} discusses systematic uncertainties. Section~\ref{sec:results} presents the search results and lifetime-limit calculation, and Section~\ref{sec:conclusion} summarizes the conclusions.

\section{The Super-Kamiokande Detector}
\label{sec:detector}

Super-Kamiokande (Super-K, SK) is a large cylindrical water-Cherenkov detector located 1 km underground (2700 meters water equivalent) in the Kamioka mine in Gifu, Japan. 
Full details of the detector can be found in Refs.~\cite{fukuda2003, mine2024}, but here we summarize only the features most relevant to the present analysis.

The SK detector consists of two optically separated regions: the inner detector (hereafter ``ID'') and the outer detector (hereafter ``OD''). 
The ID contains 32 kilotons of ultrapure water, corresponding to approximately $1.07 \times 10^{33}$ oxygen nuclei, and is lined with 
approximately 11,100 inward-facing 20-inch photomultiplier tubes (hereafter ``PMTs'') that detect Cherenkov light emitted by charged particles. 
Under normal operating conditions the photocathode coverage of the ID is approximately 40\%.
However, during the SK-II period (2002-2005) the number of PMTs was reduced to ~5,200 following an accident in 2001 that destroyed approximately half of the original PMTs. Accordingly, SK-II was operated with only 19\% coverage.
On the other hand, the OD consists of 1,885 outward-facing 8-inch PMTs and wavelength-shifting plates mounted on the ID support structure.
It's walls are lined with reflective Tyvek\textregistered{} to enhance light collection and allow for more efficient rejection of incoming cosmic ray 
muons and for identification of particles that exit the ID. 

SK collected data over multiple data-taking periods, beginning in 1996. 
For this analysis, we utilized data from all five pure-water phases, as summarized in Table~\ref{tab:sk_phases}. 
The first three periods from 1996 to 2008, SK-I, -II, and -III, differed from later periods primarily in their readout electronics.
Events were recorded within a fixed event gate of about 1.3 $\mu$s ($-300$ to $+1000$ ns around the trigger), which introduced a dead time of roughly 800--1200 ns after a primary event and limited the efficiency for tagging Michel electrons shortly afterward; an impedance mismatch between the electronics and PMT cables produced signal reflections in a similar time window but was a sub-dominant effect. In 2008, corresponding to SK-IV and later, an electronics upgrade introduced a much wider 40 $\mu$s gate ($-5$ to $+35$ $\mu$s around the trigger) for high-energy events and enabled essentially dead-time-free recording of all PMT hits (apart from a residual $\sim$1 $\mu$s channel dead time in the QTC), substantially improving Michel-electron tagging.
During these periods events were formed using offline software trigger algorithms.
For the purposes of this analysis, the difference between the first three and last two periods was a  
higher Michel electron tagging efficiency in the latter.
Otherwise the detector performance was comparable across the data sets.
More details of the detector and its calibration procedures, including the recent detector calibrations, can be found in Refs.~\cite{fukuda2003, yamada2010, nishino2009, abe2014calib, mine2024}.

\begin{table}[htbp]
\caption{Summary of the SK pure-water phases used in this analysis. The total livetime is 6511.3 days (17.827 years, using 365.25 days per year). With a 22.5-kiloton fiducial mass, the total exposure is 0.4011 megaton-years.}
\label{tab:sk_phases}
\begin{tabular}{lccc}
\hline\hline
Phase & Dates & Livetime (Days) & Photocoverage (\%) \\
\hline
SK I   & 1996--2001 & 1489.2 & 40 \\
SK II  & 2002--2005 & 798.6  & 19 \\
SK III & 2006--2008 & 518.1  & 40 \\
SK IV  & 2008--2018 & 3244.4 & 40 \\
SK V   & 2019--2020 & 461.0  & 40 \\
\hline\hline
\end{tabular}
\end{table}

\section{Simulation}
\label{sec:simulation}

In the search for nucleon decay signals, atmospheric neutrino interactions constitute the primary background, particularly in the energy range of a few to several GeV. 
The dominant background processes include charged-current single pion production (CC1$\pi$)~\cite{rein1981} and deep inelastic scattering (DIS), where a neutrino can produce multiple pions in the final state.
These processes can generate event topologies with charged leptons and multiple pions that closely resemble the trinucleon decay signature, leading to background events with similar ring multiplicities and energy distributions to those expected from the signal.

To calculate the signal efficiency and estimate the expected atmospheric neutrino background in the signal sample, we generated Monte Carlo (MC) simulations for the trinucleon decay and atmospheric neutrinos. 
We produced separate MC samples for both the trinucleon decay signal and atmospheric neutrino background for each SK detector period. 
The generation and analysis of the trinucleon decay MC follow previous SK analyses of single and dinucleon decay~\cite{gustafson2015, matsumoto2022} and is described below.

\subsection{Trinucleon Decay Simulation}

Given that the decay mode involves three nucleons, we focused on decays occurring within the oxygen nucleus. 
The simulation takes into account Fermi motion, correlations between nucleons, the nuclear binding energy, as well as pion interactions within the nucleus. 
The impact of pion interaction during propagation through the dense nuclear medium is particularly significant in this study and is discussed in more detail below.

The nucleon Fermi momentum in $^{16}$O is simulated based on data from electron-$^{12}$C scattering experiments~\cite{nakamura1976}. 
When simulating a decay, the effect of the nuclear binding energy is introduced by modifying the proton mass according to, $M_{0p} = M_{p} - E_b$, where $M_{0p}$ is the modified proton mass, $M_{p}$ is the proton rest mass, and $E_b$ is the nuclear binding energy. 
Here, $E_b$ is sampled from a Gaussian distribution with a mean of 39.0 MeV (standard deviation of 10.2 MeV) for S-state protons, and a mean of 15.5 MeV (standard deviation of 3.8 MeV) for P-state protons. 
The ratio of S-state to P-state protons is taken as 1:3, based on the nuclear shell model~\cite{mayer1955}.

During the trinucleon decay process, the kinematics of the protons may be distorted due to binding effects with surrounding nucleons, a phenomenon referred to as correlated decay. 
The probability of such correlated decays is estimated to be about 10\% for each nucleon pair~\cite{yamazaki1999}. Since three protons are involved, there are three possible correlated pairs; each pair is treated independently with a 10\% correlation probability. The effect of correlated decay on the signal efficiency is evaluated as a systematic uncertainty by varying the correlation probability between 0\% and 20\%.
Additionally, the positions of the decaying nucleons within the $^{16}$O nucleus are approximated using the Woods-Saxon nuclear density model~\cite{woods1954}. 
The decay of three nucleons leaves the residual nucleus in an excited state, which subsequently de-excites via the emission of gamma rays. 
Individual de-excitation gamma rays typically have energies less than 10 MeV, producing a total energy deposit less than ~30 MeV.
Typically this light is difficult to distinguish from the light produced by other primary particles in the event.
The decay products are generated with energy and momentum uniformly distributed within the available phase space, following the approach of previous SK nucleon decay searches~\cite{Shiozawa1999}.
Pions from the decay, for example, have momenta ranging from about 200 MeV/$c$ to several GeV/$c$, with a 200 MeV/$c$ pion producing Cherenkov light comparable to that of a 30 MeV electron. 
The positron also has a similar momentum, contributing significantly more light than the de-excitation gamma rays.

\begin{figure}[htbp]
\includegraphics[width=0.8\linewidth]{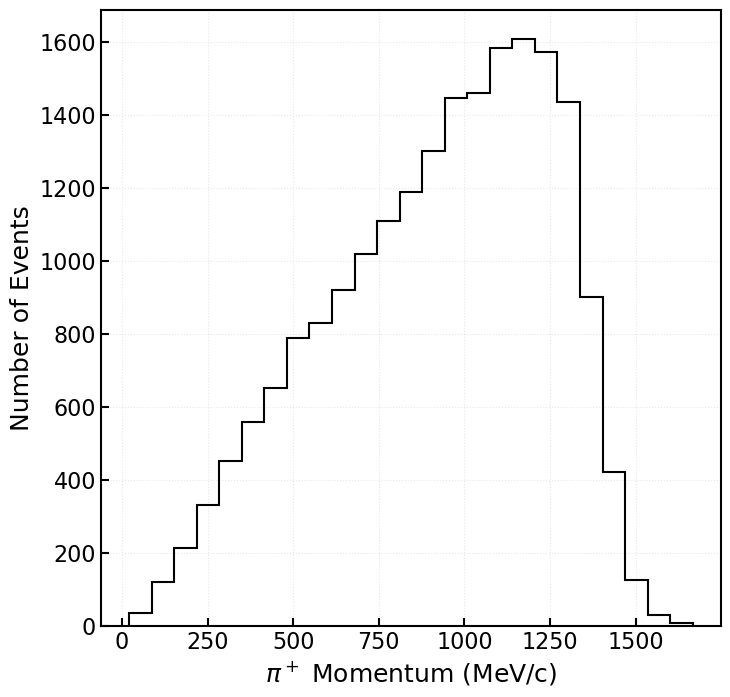}
\caption{Initial momentum distribution of $\pi^+$ mesons from trinucleon decay before FSI.}
\label{fig:pion_momentum_dist}
\end{figure}

\begin{figure}[htbp]
\includegraphics[width=0.8\linewidth]{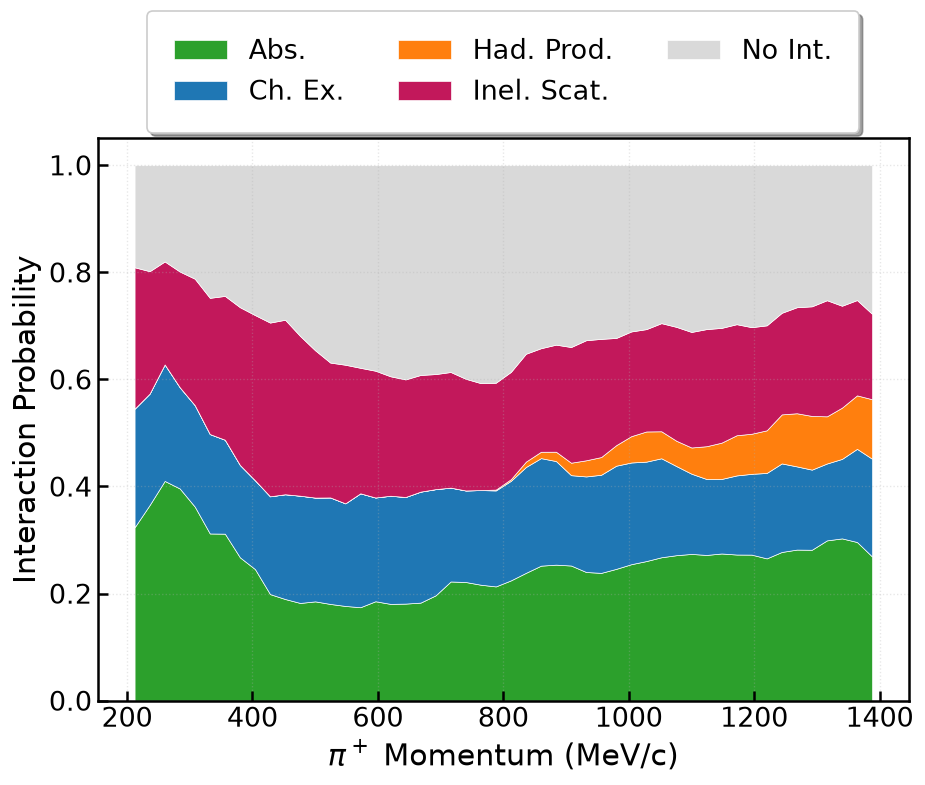}
\caption{Momentum dependence of pion FSI processes in oxygen nuclei. The stacked probability distributions show the relative contributions from different interaction channels as a function of the true $\pi^+$ momentum: absorption (Abs., green), charge exchange (Ch. Ex., blue), hadron production (Had. Prod., orange), inelastic scattering (Inel. Scat., magenta), and no interaction (No Int., gray).}
\label{fig:fsi_fraction}
\end{figure}

Pion-nucleon interactions within the nucleus (pion FSI) can have significant impact on the observed final state and are simulated using the NEUT cascade model~\cite{hayato2002}. 
Figure~\ref{fig:pion_momentum_dist} shows the initial generated momentum distribution of $\pi^+$ mesons from trinucleon decay prior to FSI, exhibiting a broad spectrum with a peak around 1200 MeV/c. 
The interaction probabilities for different FSI processes exhibit a strong momentum dependence, as illustrated in Figure~\ref{fig:fsi_fraction}. 
At low momenta (below 400 MeV/$c$), inelastic scattering and absorption dominate the interaction landscape. 
The inelastic channel maintains a significant probability across all momentum ranges, accounting for approximately 20\% of the interactions. 
Charge exchange processes, while maintaining a relatively constant probability of around 10--20\% across the momentum range, play a crucial role in modifying the event topology by converting charged pions to neutral ones. 
The probability of pions traversing the nuclear medium without interaction ranges from about 20\% at 200 MeV/$c$ to approximately 35\% at 800 MeV/$c$. 
Above this threshold this survival probability decreases again as hadron production channels are opened.

The effects of these FSI on the number of $\pi^+$ and $\pi^0$ mesons from $^{16}\text{O}(ppp) \rightarrow ^{13}\text{C} \pi^+ \pi^+ e^+$ decay are summarized in Table~\ref{tab:fsi_impact}. 
After FSI the fraction of events with the same number of pions as in the initial state is estimated to be 47.8\%. 
The next most probable combination, occurring in 18.1\% of events, consists of one $\pi^+$ and one $\pi^0$. 
This combination is typically realized when a $\pi^+$ exchanges charge with a neutron and produces a $\pi^0$. 
In some cases, this could also happen when one $\pi^+$ is absorbed while the other $\pi^+$ interacts with a nucleon and generates an additional $\pi^0$. 
Other combinations, including those involving multiple charge exchange or absorption processes, occur with lower probabilities.

Once the positron and pions exit the nucleus, their propagation and Cherenkov light emission are simulated using a custom detector model based on GEANT3~\cite{cern1994}. 
Charged pions are subject to further hadronic interactions in water, while neutral pions decay before such interactions can occur. 
The propagation of charged pions at momenta below 500~MeV$/c$ is simulated using the nuclear effect part of NEUT~\cite{nakahata1986}, while higher momentum pions are handled by GCALOR~\cite{albanese1980}.

MC events are generated within 1 meter from the wall of the ID, expanding the fiducial volume to accommodate the requirements of machine learning models used later in the analysis to identify common features of trinucleon decay events and to distinguish them from backgrounds. 
For each SK period, a total of 100,000 signal events were generated.

\begin{table}[htbp]
    \centering
    \caption{Impact of FSI on the number of $\pi^+$ and $\pi^0$ mesons in the decay $^{16}\text{O}(ppp) \rightarrow ^{13}\text{C} \pi^+ \pi^+ e^+$. Values shown are percentages of the total sample. The original signal combinations (two $\pi^+$ with no $\pi^0$, or one $\pi^+$ with one $\pi^0$ after charge exchange) are highlighted with bold text. Due to strong FSI within the oxygen nucleus, only about 66\% of events maintain these viable signal combinations, with the remaining events significantly altered through various nuclear interactions. $N_{\pi^0}$ denotes the number of $\pi^0$ mesons.}
    \label{tab:fsi_impact}
    \vspace{0.2cm}
    \begin{tabular}{ccccc}
        \toprule
        \textbf{$N_{\pi^0}$} & \textbf{0 $\pi^+$} & \textbf{1 $\pi^+$} & \textbf{2 $\pi^+$} & \textbf{3 $\pi^+$} \\
        \midrule
        0 & 0.1\% & 12.1\% & \textbf{47.8\%} & 4.3\% \\
        1 & 2.8\% & \textbf{18.1\%} & 6.7\% & 1.9\% \\
        2 & 2.0\% & 2.6\% & 1.4\% & 0.4\% \\
        \bottomrule
    \end{tabular}
\end{table}

\subsection{Atmospheric Neutrinos}

Atmospheric neutrinos are the primary background to this study. 
For this analysis, we generate atmospheric neutrino MC samples for each SK period using the Honda atmospheric neutrino flux model~\cite{honda2011} and the NEUT neutrino interaction generator~\cite{hayato2002}. 
These samples include neutrino interactions in water, where charged-current single pion production (CC1$\pi$) and deep inelastic scattering (DIS) constitute the dominant background processes. 
Both interaction modes can produce final states that closely mimic our signal topology: 
CC1$\pi$ events can generate combinations with multiple pions and charged leptons through FSI, while DIS processes can directly produce multi-pion final states with kinematics similar to our trinucleon decay signal.

Charged-current single pion production (CC1$\pi$) follows the reaction:

\[
\nu + N \rightarrow l + N' + \pi
\]

\noindent where $\nu$ represents the neutrino, $N$ is the initial nucleon, $l$ is the charged lepton, $N'$ is the final-state nucleon, and $\pi$ is the produced pion.

Deep inelastic scattering (DIS) can be further categorized into charged-current (CCDIS) and neutral-current (NCDIS) interactions, represented by the following reactions:

\[
\text{CCDIS: } \nu + N \rightarrow l + N' + \text{hadrons}
\]
\[
\text{NCDIS: } \nu + N \rightarrow \nu + N' + \text{hadrons}
\]

In DIS processes, the final-state products can include multiple pions, as well as other hadrons such as $\eta$ mesons and kaons.
In NEUT, the hadronic system is modeled differently depending on the invariant hadronic mass $W$. 
For $1.3\,\text{GeV}/c^2 \leq W \leq 2.0\,\text{GeV}/c^2$, only multiple-pion production is simulated with multiplicities based on bubble chamber experimental data. 
For $W > 2.0\,\text{GeV}/c^2$, the hadronic final states are modeled using PYTHIA/JETSET, allowing for production of additional mesons such as kaons and $\eta$ mesons. 
The nucleon's internal structure is described using GRV98 parton distribution functions~\cite{gluck1998} with Bodek-Yang corrections~\cite{yang1999,bodek2003} applied in the low $Q^2$ region.
These atmospheric neutrino events were simulated using the same detector simulation program as the signal events, incorporating GEANT3 with GCALOR for hadron interactions in water and NEUT for low-momentum pion interactions, following the standard SK simulation framework~\cite{fukuda2003, yamada2010}.

The atmospheric neutrino MC samples were generated with statistics equivalent to a 500-year exposure of each SK period.
Background estimates were reweighted using three-flavor neutrino oscillation probabilities applied using the PDG 2022 global best-fit parameters: $\sin^2\theta_{12} = 0.307$, $\sin^2\theta_{13} = 0.0220$, $\sin^2\theta_{23} = 0.546$, $\Delta m^2_{21} = 7.53 \times 10^{-5}$~eV$^2$, $\Delta m^2_{32} = 2.453 \times 10^{-3}$~eV$^2$, and $\delta_{CP} = 0$. This approach is consistent with that used in Ref.~\cite{jung2025}.

\section{Analysis}
\label{sec:analysis}

In this study, we first applied event selections using the standard SK reconstruction algorithm (APFit)~\cite{Shiozawa1999, ashie2005, mine2024} to identify events broadly consistent with the expected trinucleon decay mode.
These pre-selected events are then subjected to advanced machine learning (ML) algorithms for further classification. 

\subsection{Preliminary Event Selection}

For this analysis, we first reconstructed the interaction vertex using timing and charge information from the PMTs to identify events within the SK detector. 
We then selected Fully Contained Fiducial Volume (hereafter ``FCFV'') events, which satisfy two critical criteria: all charged particles must be confined within the ID with no significant activity in the OD, and 
the reconstructed vertex must lie within the fiducial volume, defined as the region 2 meters inward from the ID walls. 
Additionally events are required to have at least 30~MeV of visible energy deposited in the detector.

Events passing the FCFV selection underwent a reconstruction process to determine key parameters, including vertex position, number of Cherenkov rings, direction, momentum, and particle type. 
The reconstruction algorithm first determines the vertex position using timing and charge information from the PMTs, then identifies Cherenkov rings using the Hough transformation method~\cite{davies1997machinevision, Shiozawa1999}.
Particle identification is performed by analyzing the photon charge distribution within each ring and each ring in an event is identified as either showering, representative of an electron or photon,
or non-showering, representing a muon or charged pion.
The corresponding momentum is reconstructed from the total charge associated with each Cherenkov ring.
Michel electrons from stopping muons are identified by detecting clusters of PMT hits within a 20-microsecond window following the primary event. 
Full details of this reconstruction process, including the algorithm refinements introduced after SK-I such as the ring correction, are available in Refs.~\cite{Shiozawa1999, ashie2005, mine2024}.

The nominal target event topology for this search includes two $\pi^+$ and one $e^+$, which should produce two non-showering and one showering ring. 
Often the $\pi^+$ will decay within the inner detector producing a Michel electron that can be tagged and used to improve signal identification.
However, in addition to the FSI described above, pions emerging from the oxygen nucleus can undergo secondary interactions (SI) 
with nuclei in the water, undergoing charge exchange, absorption, or additional particle production, altering the final event topology.
The pre-selection described below has been developed to include many signal events even after accounting for these processes:

\begin{itemize}
    \item Require FCFV: Event vertices must be greater than 2~m from the nearest ID wall with limited OD activity. Further, events must have more than 30~MeV of visible energy.
    
    \item Require events with three or four Cherenkov rings: Signal events are expected to have at least three rings (two from $\pi^+$ and one from the primary $e^+$), with a fourth ring possible if a $\pi^+$ scatters at a large angle or undergoes charge exchange to produce a $\pi^0$ that promptly decays to two photons. We note that  Particle identification (PID) to distinguish showering and non-showering rings was not applied at this stage to preserve events for machine learning training.
    
    \item Require one or two Michel electrons: This targets signal events where at least one $\pi^+$ successfully completes its decay chain ($\pi^+ \to \mu^+ \to e^+$).

    \item Require total visible energy below 3 GeV: This cut, corresponding to the mass of three nucleons (approximately 3 GeV), retains signal events while rejecting higher-energy background neutrino interactions. No further lower energy limit was set to maximize the event sample available for subsequent machine learning analysis.
\end{itemize}

\begin{figure}[htbp]
    \centering
    \includegraphics[width=0.8\linewidth]{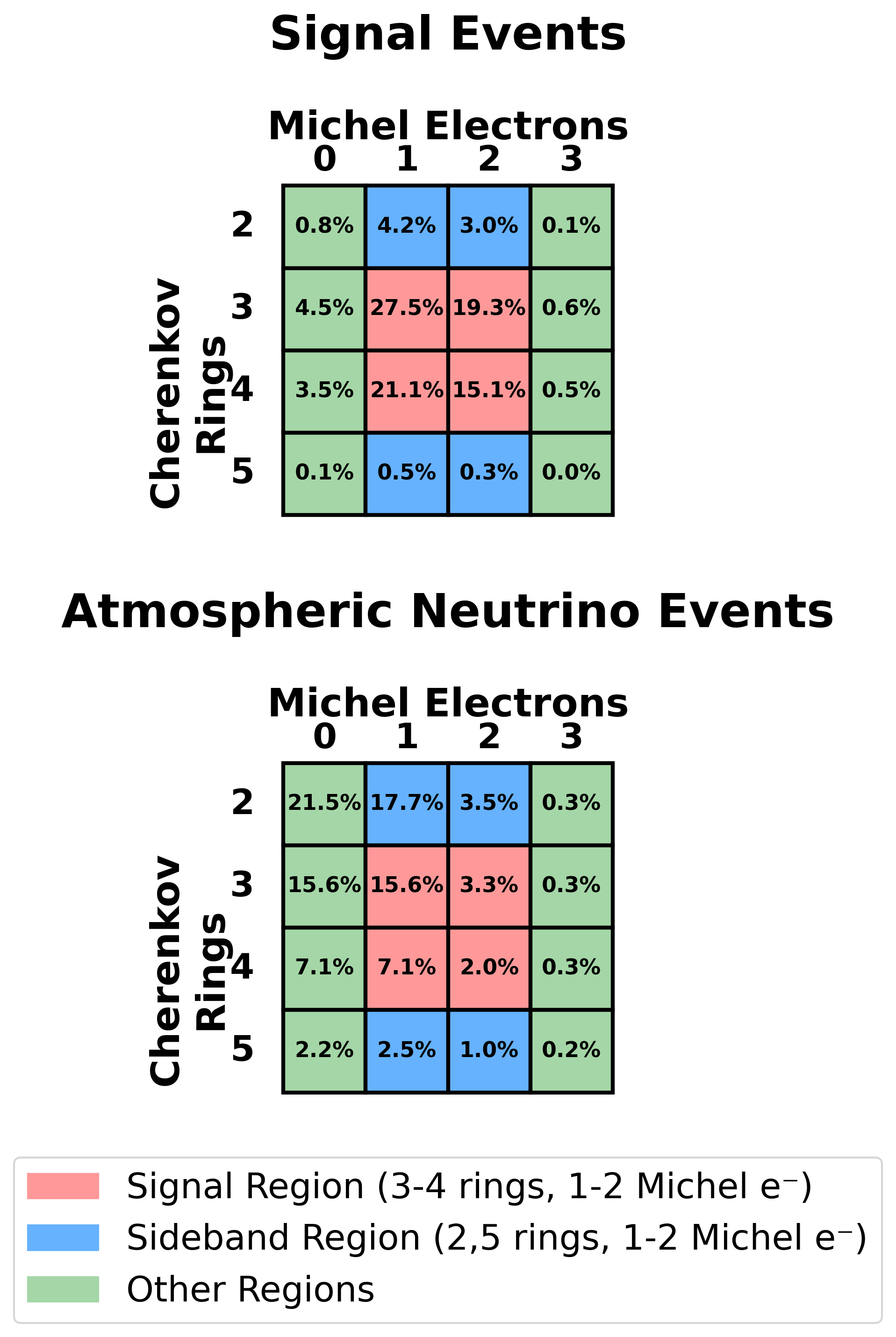}
    \caption{Event topology distributions after FSI and SI, categorized by number of Cherenkov rings and Michel electrons for both signal events (top) and atmospheric neutrino events (bottom). The signal region (3-4 rings, 1-2 Michel electrons) is highlighted in red, while the sideband region (2 or 5 rings, 1-2 Michel electrons) used for CNN systematic uncertainty assessment is highlighted in blue. Other event topologies are shown in green.}
    \label{fig:event_topology_distribution}
\end{figure}

Figure~\ref{fig:event_topology_distribution} presents the distribution of events across the different topologies after the FCFV cut (including the fiducial volume and visible energy requirements) but before applying the other selection criteria.
The pre-selection's signal region is shown (highlighted in red) together with sidebands discussed below (blue) and other events (green).
Trinucleon decay events pass the preselection with 82.9\% efficiency, whereas only 28.0\% of atmospheric neutrino events survive. 

\subsection{Machine Learning Methods}
\label{sec:ml_analysis}

Our analysis employs machine learning techniques that process detector response patterns to identify characteristic signatures of trinucleon decay events. This is the first time a CNN has been applied to a nucleon decay search at SK. The machine learning approach systematically analyzes spatial correlations in the charge distribution patterns recorded by photomultiplier tubes to distinguish signal events from background.

The fundamental computational unit in this analysis is a convolution neural network, termed a CNN, which divides the detector response into discrete spatial regions and analyzes local correlations through a series of mathematical transformations. These transformations are applied sequentially, with each subsequent operation building upon the results of previous calculations to identify increasingly complex spatial patterns in the detector response. The method's efficacy stems from its ability to recognize subtle correlations in detector response patterns that may not be apparent in conventional reconstruction approaches.

CNNs excel at recognizing complex structures within Cherenkov rings by automatically extracting significant features from the input images, such as ring shapes and intensity patterns. This enables more precise classification of high-momentum pion events. The effectiveness of this approach builds on foundational studies, such as the CNN architecture proposed by LeCun et al.~\cite{lecun1998gradient}.

To address the impact of hadron production for pion momenta above 500 MeV/c, we employed CNNs for feature extraction. 

\subsubsection{Unfolding the SK Detector}

In our study, the SK detector, a large cylindrical structure equipped with more than ten thousand PMTs, was used to record physical events. To effectively apply CNNs for event analysis, we transformed the surface images of the cylindrical detector into a 2D plane. For this transformation, we used the total charge collected by each PMT throughout the entire event duration (1.3 $\mu$s for SK-I to SK-III, and 40 $\mu$s for SK-IV and SK-V), which includes charges from both the primary interaction and any subsequent decay processes. The timing information of PMT hits was not included in the analysis to maintain simplicity and focus on the spatial charge distribution.

\begin{figure}[ht]
    \centering
    \includegraphics[width=\columnwidth]{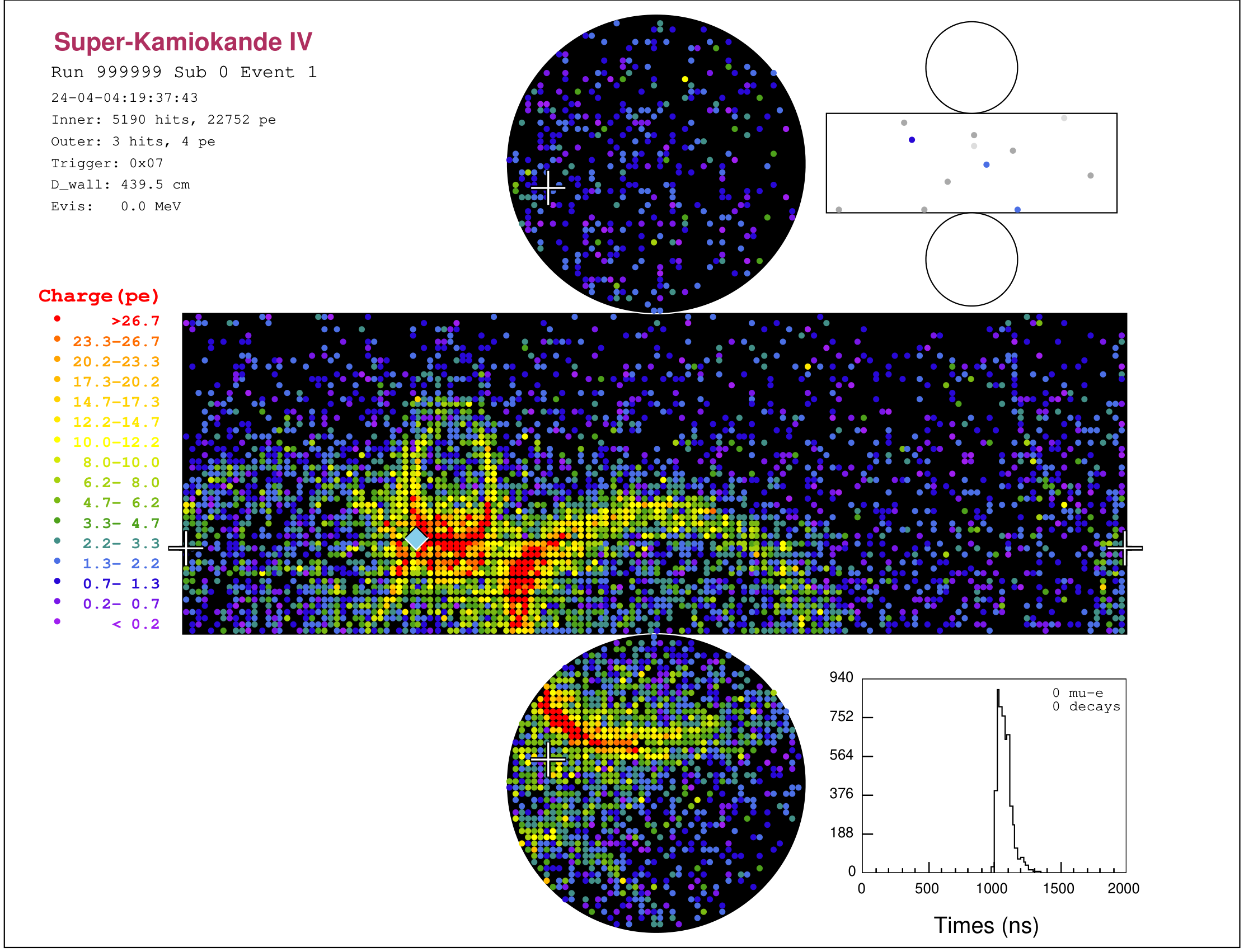}\\[4pt]
    \includegraphics[width=\columnwidth]{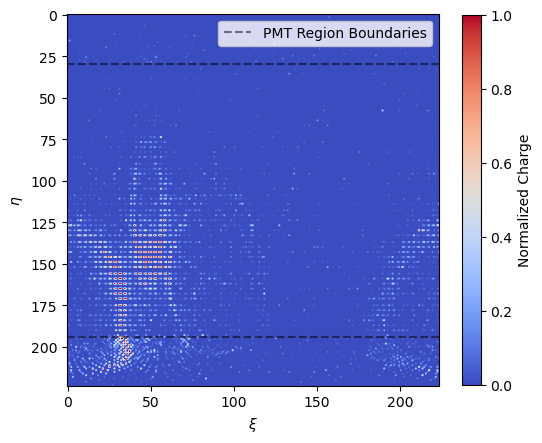}
    \caption{MC simulated images of trinucleon decay and their corresponding Mercator projections. The upper image shows the original cylindrical detector simulation, while the lower image demonstrates the 2D plane obtained using our Mercator projection method.}
    \label{fig:mercator_projection}
\end{figure}

To facilitate the application of CNN analysis, we developed a mapping procedure to transform the cylindrical detector surface into a rectangular image representation. This transformation defines a two-dimensional coordinate system ($\xi$, $\eta$), where $\xi$ represents the azimuthal coordinate and $\eta$ denotes the vertical position in the projected plane. The transformation preserves the topological properties of Cherenkov ring patterns, which is essential for subsequent pattern recognition. Figure~\ref{fig:mercator_projection} demonstrates this transformation applied to simulated trinucleon decay events, showing both the original cylindrical detector view and the resulting 2D Mercator projection.

The mapping procedure implements distinct transformations for PMTs located on the barrel and end-caps of the detector. Each PMT maps to at most one pixel in the 224×224 grid of the projection, with unoccupied grid positions assigned zero charge values. For barrel PMTs, the azimuthal angle \(\phi\) is mapped linearly to the \(\xi\) coordinate with a constant offset (Eq.~\ref{eq:barrel_xi}):
\begin{equation}
    \xi = \left\lfloor \frac{\phi}{2\pi} \times N_{\xi} \right\rfloor + \xi_0 \label{eq:barrel_xi},
\end{equation}
where \(N_{\xi} = 224\) is the number of horizontal pixels in the output image and \(\xi_0\) is a horizontal offset parameter for proper alignment.
\noindent The vertical position \(z\) is transformed to \(\eta\) via:
\begin{equation}
    \eta = \left\lfloor \frac{z - z_{\text{min}}}{z_{\text{max}} - z_{\text{min}}} \times N_{\eta_{\text{barrel}}} \right\rfloor + \eta_0 \label{eq:barrel_eta}.
\end{equation}

\noindent For end-cap PMTs, the transformation first computes their radial coordinate \(\rho\) in the end-cap plane:
\begin{equation}
    \rho = \sqrt{x^2 + y^2} \label{eq:rho}.
\end{equation}
\noindent This radial coordinate is subsequently mapped to \(\eta\) :
\begin{equation}
    \eta = \left\lfloor \frac{\rho}{\rho_{\text{max}}} \times N_{\eta_{\text{endcap}}} \right\rfloor + \eta_0 \label{eq:endcap_eta},
\end{equation}
\noindent while the azimuthal position determines \(\xi\) through:
\begin{equation}
    \xi = \left\lfloor \frac{\arctan2(y,x)}{2\pi} \times N_{\xi} \right\rfloor + \xi_0 \label{eq:endcap_xi}.
\end{equation}

\noindent The transformation parameters are defined as follows: \(N_{\xi} = 224\) and \(N_{\eta} = 224\) specify the total dimensions of the output image, encompassing both the barrel and end-cap regions after projection into a 2D plane; \(z_{\text{min}}\) and \(z_{\text{max}}\) denote the vertical height limits of the barrel section of the detector, while \(\rho_{\text{max}}\) represents the maximum radial extent of the end-caps. The constants \(\xi_0\) and \(\eta_0\) are offsets ensuring proper alignment in the transformed coordinate system. The floor function \(\lfloor \cdot \rfloor\) ensures discrete pixel coordinates in the output image.

This cylindrical-to-planar projection introduces geometric distortions, particularly in the top and bottom (end-cap) sections of the detector compared to the barrel region. These distortions have a negligible impact on our analysis for two reasons. First, the same fixed projection is applied identically to both the simulated and the observed events, so any distortion is common to signal MC, background MC, and data, and is therefore learned consistently by the network rather than introducing a data/MC bias. Second, the fiducial-volume requirement (vertex more than 2~m from the wall) and the multi-ring topology of the signal mean that the majority of Cherenkov rings are imaged on the barrel, where the projection is close to area-preserving; events dominated by the more strongly distorted end-cap regions are a small fraction of the sample.

The geometric shape and intensity distribution of Cherenkov rings are directly related to the particle's direction and speed. Therefore, ring features contain not only the particle's momentum information but also characteristics related to the particle type. CNNs are particularly effective in processing image data, excelling at extracting geometric features such as edges and curves. By using Cherenkov rings as the primary features, CNNs can leverage their convolutional layers to incrementally extract edge information from the image, identifying the ring boundaries in earlier layers and capturing the overall shape and details in deeper layers. This layered feature extraction enables CNNs to effectively distinguish between different event types, especially in the case of multiple partially overlapping rings as expected from the trinucleon decay signal.

\subsubsection{Network Architecture}
We implemented a modified version of the MobileNetV3 architecture\cite{mobilenetv3}, which performs sequential mathematical operations on detector response data. The architecture consists of two primary computational components, designed to optimize both computational efficiency and pattern recognition capability.

First, a spatial averaging operation across detector regions is applied, followed by a weighting calculation that determines the relative importance of different spatial regions in the detector response. This component, known as the Squeeze-and-Excitation mechanism in computational literature~\cite{hu2018senet}, enables adaptive weighting of detector signals based on their spatial distribution. These weights are then applied to modulate the detector signal intensities according to their calculated significance.

Second, a series of dimensional reduction operations, referred to as the Bottleneck structure, that process the detector response data through three stages: an initial transformation that reduces the dimensionality of the signal, followed by a spatial correlation analysis using either 3×3 or 5×5 matrices, and a final transformation that restores the original signal dimensions. This structure, based on the inverted-residual bottleneck design introduced in MobileNetV2~\cite{sandler2018mobilenetv2}, allows for efficient computation while preserving the essential spatial correlations in the detector response.

The detector response data is preprocessed by projecting the cylindrical detector geometry onto a planar surface. This projection is then discretized into a numerical array intended as input for the CNN, with dimensions \([3, 224, 224]\). The final two dimensions, \(N_{\xi} = 224\) and \(N_{\eta} = 224\), specify the spatial resolution of the projected image in pixels. The first dimension represents the input channels. Since the PMT charge distribution constitutes single-channel data (analogous to a grayscale image), but the CNN architecture expects a three-channel input (akin to standard RGB images), this single channel of charge information is replicated across all three channels. Consequently, the array features three identical input channels, each containing the same PMT charge distribution.

\subsubsection{Training and Classification of the $^{16}$O(ppp) $\rightarrow$ $^{13}$C $\pi^+ \pi^+ e^+$ Search}

This analysis utilized MC from five SK operational periods. For each period, we generated 100,000 trinucleon decay signal events and 150,000 atmospheric neutrino background events that passed the event selection criteria in Section IV.A, resulting in a total of 1,250,000 events for the machine learning analysis. A separate CNN model was trained for each SK period to account for the different detector configurations and operational conditions.

To ensure the reliability of our method, we employed a cross-validation process—a statistical technique that assesses model performance by partitioning data into training and testing subsets. For each SK period, the events were randomly split into two sets: 80\% for optimizing parameters and 20\% for validating performance. This split enabled a fair assessment of the method's ability to distinguish signal from background events, with each period analyzed separately to account for variations in detector conditions.

The analysis was implemented using the PyTorch framework \cite{pytorch}. The network weights were initialized from a MobileNetV3 model pre-trained on the ImageNet-1k natural-image dataset~\cite{huggingface_mobilenetv3}. This transfer-learning initialization provides the convolutional filters with generic low-level feature detectors (edges, curves, and intensity gradients) that are also relevant for Cherenkov-ring patterns. Starting from these pre-trained weights, rather than from random initialization, accelerates convergence and improves stability of the training given our limited number of simulated events, while the subsequent training on SK MC adapts the higher-level features to the detector-specific topologies.
A CNN was employed to classify the pre-selected samples, leveraging its ability to process the projected detector response data (Section IV.B.1). To optimize the CNN settings, we used a binary cross-entropy loss function—a mathematical measure that quantifies the difference between predicted and actual event classifications—to guide adjustments. Through iterative optimization, we fine-tuned key training parameters: the batch size (number of events processed simultaneously, set to 32), the regularization coefficient (parameter controlling model complexity to prevent overfitting, set to \(1 \times 10^{-4}\)), the learning rate (step size for parameter updates, set to \(1 \times 10^{-3}\)), and the number of epochs (complete passes through the training dataset, set to 30). The trained CNN model was then applied to both the simulation and observed data, producing a distribution with two distinct peaks---one corresponding to the trinucleon decay signal and the other to the atmospheric neutrino background---as shown in Figure \ref{fig:CNNOutputLog}.

\subsection{Application and Performance of the CNN in the $^{16}$O(ppp) $\rightarrow$ $^{13}$C $\pi^+ \pi^+ e^+$ Search}

The trained CNN model was applied to the pre-selected samples from all five SK periods, producing a distribution of output scores that separated trinucleon decay signal events from atmospheric neutrino background events. To determine the optimal threshold for event classification, we defined a scoring function that balances the true positive rate (TPR, fraction of correctly identified signal events) with the true negative rate (TNR, fraction of correctly identified background events):
\[
\text{Score} = \frac{\text{TPR}}{\sqrt{\text{TPR} + \text{TNR}}}.
\]

\begin{figure}[htbp]
    \centering
    \includegraphics[width=0.95\linewidth]{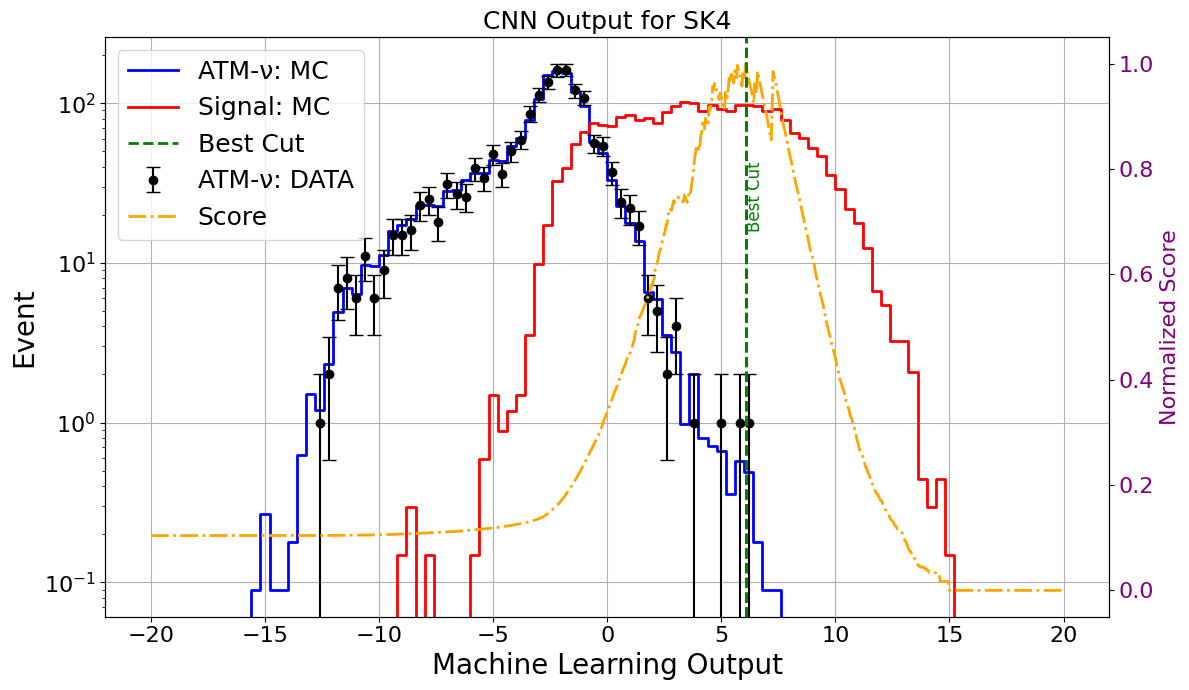}
    \caption{
        Machine learning output distribution for SK-IV period on a logarithmic scale. The red line shows the trinucleon decay signal MC events, while the blue line represents the atmospheric neutrino MC background events. Black dots with error bars indicate the experimental data for atmospheric neutrino events. The orange dash-dotted line shows the normalized score curve (right y-axis), and the green dashed vertical line indicates the optimal cut value. The separation between signal and background distributions demonstrates the CNN's discrimination power. All other SK periods show similarly good agreement between the data and the background MC; the corresponding distributions are shown in the Appendix.
    }
    \label{fig:CNNOutputLog}
\end{figure}

\begin{figure}[htbp]
    \centering
    \includegraphics[width=0.8\linewidth]{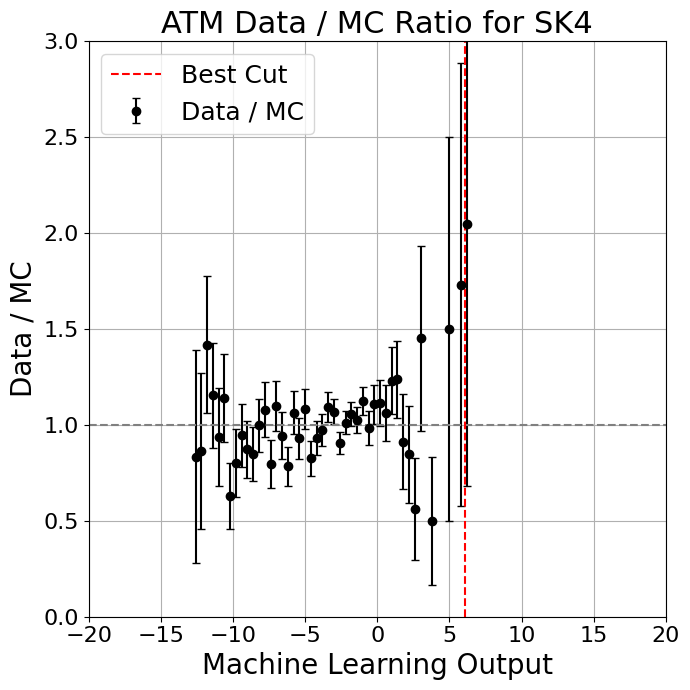}
    \caption{
        Atmospheric-neutrino data divided by atmospheric-neutrino MC as a function of the CNN output for SK-IV. Black points show the Data/MC ratio with statistical error bars, the horizontal gray dashed line marks unity, and the red dashed vertical line indicates the optimized cut. The corresponding Data/MC ratios for the other SK periods are shown in the Appendix and exhibit similarly good agreement between data and background MC.
    }
    \label{fig:CNNOutputRatio}
\end{figure}

Figure \ref{fig:CNNOutputLog} illustrates the CNN model output distribution for the SK-IV dataset, displaying both MC simulations and experimental data on a logarithmic scale. The red line represents the trinucleon decay signal MC events, while the blue line shows the atmospheric neutrino MC background events. Black dots with error bars indicate the actual experimental data, demonstrating good agreement with the atm-nu MC prediction. The orange dash-dotted line represents the normalized scoring function \( \text{Score} = \text{TPR} / \sqrt{\text{TPR} + \text{TNR}} \) (plotted against the right y-axis), which is designed to balance signal efficiency (TPR) against background rejection (related to TNR). 
The optimal cut, indicated by the green dashed vertical line, is chosen at the machine learning output value where this Score function reaches its maximum. This selection strategy aims to effectively distinguish signal from background. The clear separation between signal and background distributions, particularly visible between machine learning output values of 0 to 10, indicates the CNN's strong discriminatory power for trinucleon decay event identification. To further illustrate the discrimination capability, Figure~\ref{fig:CNNOutputRatio} shows the Data/MC ratio, providing a complementary view of the agreement between experimental data and MC predictions across different machine learning output values.

The optimal cut value is determined independently for each SK data-taking period (SK-I through SK-V). 
For each period, only events with machine learning output values greater than their respective optimized cut will be selected as the final sample. This period-specific optimization ensures consistent signal selection efficiency while maintaining effective background rejection across all operational phases (Table~\ref{tab:eff_bkg_data_transposed}). Using these optimized cuts, we can estimate both the method's efficiency and the atmospheric neutrino content in the final sample for each period.

\begin{table}[htbp]
  \centering
  \caption{Efficiencies and background estimates across different SK periods 
            for the $ppp \rightarrow \pi^+ \pi^+ e^+$ search. 
            The expected-background counts are calculated from the displayed event rates multiplied by each period's exposure, $E_i=(22.5/1000)(T_i/365.25)$ megaton-years, using the livetimes $T_i$ in Table~\ref{tab:sk_phases}. Counts are approximate because the input rates are rounded.}
  \setlength{\tabcolsep}{1.2pt}
  \begin{tabular}{lcccc}
      \hline
      & \textbf{Eff. (\%)} & \textbf{Expected} & \textbf{Event Rate} & \textbf{Data} \\ 
      & & \textbf{bkg events} & \textbf{(events/MT$\cdot$yr)} & \textbf{Events} \\ \hline
      \textbf{SK-I} \\
      \quad PreCut
        & 81.4 & 386.54  & 4213.6 & --  \\
      \quad ML Cut
        & 25.9 & 0.29   & 3.2    & 0  \\ \hline
      \textbf{SK-II} \\
      \quad PreCut
        & 80.9 & 211.05  & 4290.1 & --  \\
      \quad ML Cut
        & 22.8 & 0.21   & 4.3    & 0  \\ \hline
      \textbf{SK-III} \\
      \quad PreCut
        & 80.4 & 135.88  & 4257.3 & --  \\
      \quad ML Cut
        & 25.7 & 0.10   & 3.1    & 0  \\ \hline
      \textbf{SK-IV} \\
      \quad PreCut
        & 81.3 & 849.65 & 4251.2 & --  \\
      \quad ML Cut
        & 25.8 & 0.64    & 3.2    & 1  \\ \hline
      \textbf{SK-V} \\
      \quad PreCut
        & 81.2 & 121.15  & 4266.0 & --  \\
      \quad ML Cut
        & 25.9 & 0.09    & 3.0    & 0  \\ \hline
  \end{tabular}
  \label{tab:eff_bkg_data_transposed}
\end{table}

In our analysis of Cherenkov ring concentration, we observe that trinucleon decay events tend to produce more dispersed ring distributions, whereas atmospheric neutrino events are characterized by a higher spatial concentration of rings. 
This concentration is quantified by the mean angle between the highest energy ring and the other rings, reflecting their relative positions within the detector. For trinucleon decay events, low initial momenta of the decaying nucleons results in decay products being more isotropically distributed.

In contrast, atmospheric neutrino events, particularly those involving deep inelastic scattering (DIS), typically exhibit smaller mean angles because the incoming neutrino's directional momentum tends to produce more concentrated ring patterns along the direction of the incoming neutrino. However, this directionality is partially altered by FSI within the nuclear medium, which can significantly alter the trajectories of the outgoing particles and lead to more diffuse angular distributions than would be expected from pure DIS kinematics alone.

Though we performed comparative analyses across all SK periods (SK-I to SK-V), here we focus on SK-II and SK-IV only.
The results from SK-I, SK-III, and SK-V showed patterns similar to SK-IV, without significant deviations and SK-II has features distinct from the other periods.

\begin{figure*}[!t]
    \centering
    \includegraphics[width=0.47\textwidth]{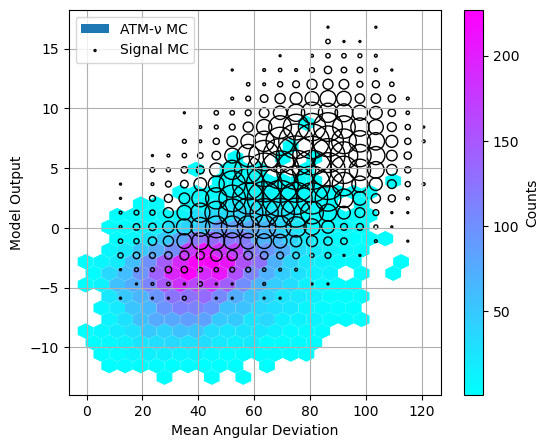}\hfill
    \includegraphics[width=0.47\textwidth]{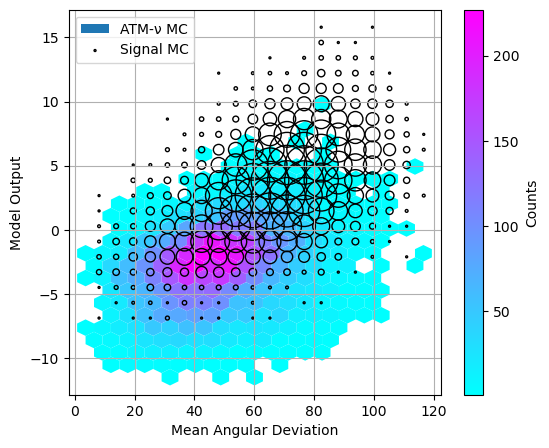}
    \caption{CNN model output versus the mean angular deviation of the final-state particles (computed from generator-level directions after FSI) for SK-IV (left) and SK-II (right). The colored hexagonal bins show the atmospheric-neutrino background MC density (color scale at right), and the open black circles show the trinucleon decay signal MC, with the marker area proportional to the number of events. The mean angular deviation, a generator-level physics quantity, is consistent between the two periods, but the signal and background populations are clearly separated along the CNN output axis in SK-IV while overlapping substantially more in SK-II. This reduced separation reflects the lower PMT coverage of SK-II (roughly half that of the other periods), which makes the Cherenkov rings less sharply defined and thereby degrades the network's discriminating power.}
    \label{fig:ConcentrationComparison}
\end{figure*}

Figure \ref{fig:ConcentrationComparison} illustrates the distribution of the mean angular deviation for the SK-IV period (left panel) and the SK-II period (right panel). The mean angular deviation is calculated using the true (generator-level) directions of the final-state particles from the MC simulation, taken after FSI. It is computed as:

\[
\theta_{\text{avg}} = \frac{1}{N-1} \sum_{i=1}^{N-1} \theta_i,
\]

\noindent where \(\theta_i\) is the angular difference (in radians) between the direction of the highest-energy final-state particle and the direction of the \(i\)-th particle, and \(N\) is the total number of final-state particles in the event (excluding the highest-energy particle). Because this quantity is computed from generator-level directions after FSI, it is a property of the underlying interaction physics and is therefore essentially the same for a given event sample across SK periods; in particular, the atmospheric-neutrino background distribution in mean angular deviation is consistent between SK-II and SK-IV. The pronounced difference between the two periods instead appears along the CNN output axis: in SK-IV the signal and background populations are well separated in the model output, whereas in SK-II they overlap considerably more. This degraded separation arises from the different PMT configuration of SK-II, which operated with roughly half the photocathode coverage of the other periods. With fewer active PMTs collecting Cherenkov photons, the ring patterns are less sharply defined in the detector images, which reduces the discriminating power of the CNN for SK-II rather than shifting the underlying angular-deviation distribution. The marked difference between the SK-II and SK-IV results thus reflects the direct impact of PMT coverage on the clarity of the Cherenkov-ring images and hence on the network's classification performance.

\begin{figure*}[!t]
  \centering
  \includegraphics[width=0.85\textwidth]{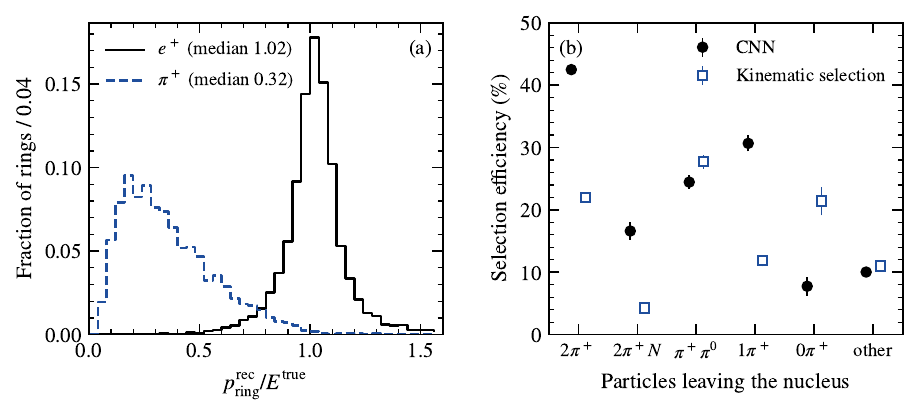}
  \caption{SK-IV signal MC. (a) Ratio of the reconstructed ring momentum to the true energy for APFit rings matched to the true $e^+$ (solid) and $\pi^+$ (dashed) leaving the nucleus. (b) Selection efficiency of the CNN (filled circles) and the kinematic selection (open squares) for each class of particles leaving the nucleus after FSI: two $\pi^+$ without ($2\pi^+$) and with ($2\pi^+N$) knocked-out nucleons, one $\pi^+$ and one $\pi^0$ ($\pi^+\pi^0$), one $\pi^+$ ($1\pi^+$), no $\pi^+$ ($0\pi^+$), and other final states.}
  \label{fig:ring_eff}
\end{figure*}

\subsection{Comparison with a Conventional Kinematic Selection}
\label{sec:kinematic_comparison}

To understand which signal events the CNN selects, we compared it event by event with a conventional selection based on APFit kinematics, similar to those used in previous SK nucleon-decay searches~\cite{Shiozawa1999, ashie2005, mine2024}. This kinematic selection requires two to four rings, fewer than three Michel electrons, $1600<M_{\rm inv}<3000$~MeV/$c^2$, and $P_{\rm tot}<900$~MeV/$c$, where $M_{\rm inv}$ and $P_{\rm tot}$ are the invariant mass and total momentum reconstructed from all rings. Both selections are applied to the same 8,589 SK-IV signal-MC events that pass the precut. For each event we use the full simulation history: the particles leaving the nucleus after FSI, as given by NEUT, and the subsequent history of each $\pi^+$ in water, as given by the detector simulation. Reconstructed rings are matched to the true $e^+$ and $\pi^+$ leaving the nucleus by direction, requiring an opening angle below $15^\circ$.

\begin{table}[tb]
  \centering
  \caption{Comparison of the CNN and the kinematic selection for SK-IV. Efficiencies and expected backgrounds are computed as in Table~\ref{tab:eff_bkg_data_transposed}; the background of the kinematic selection is scaled from that of the CNN by the ratio of atmospheric-neutrino MC events retained by the two selections in the same MC sample (15 and 5 events). The lower rows give the overlap in the 8,589 signal-MC events after the precut.}
  \label{tab:kinematic_comparison}
  \begin{tabular}{lcc}
    \toprule
     & \textbf{CNN} & \textbf{Kinematic} \\
    \midrule
    Signal efficiency (\%) & 25.8 & 16.2 \\
    Expected background events & 0.64 & $\approx 1.9$ \\
    \midrule
    Signal-MC events selected & 2,677 & 1,714 \\
    \quad selected by both & \multicolumn{2}{c}{792} \\
    \quad fraction selected by both & 30\% & 46\% \\
    \quad selected by this method only & 1,885 & 922 \\
    \bottomrule
  \end{tabular}
\end{table}

Table~\ref{tab:kinematic_comparison} summarizes the result. Only 792 events are selected by both methods, 30\% of the CNN sample and 46\% of the kinematic sample, so neither selection contains the other. The two exclusive populations have distinct physical origins.

\begin{figure*}[!t]
  \centering
  \includegraphics[width=0.85\textwidth]{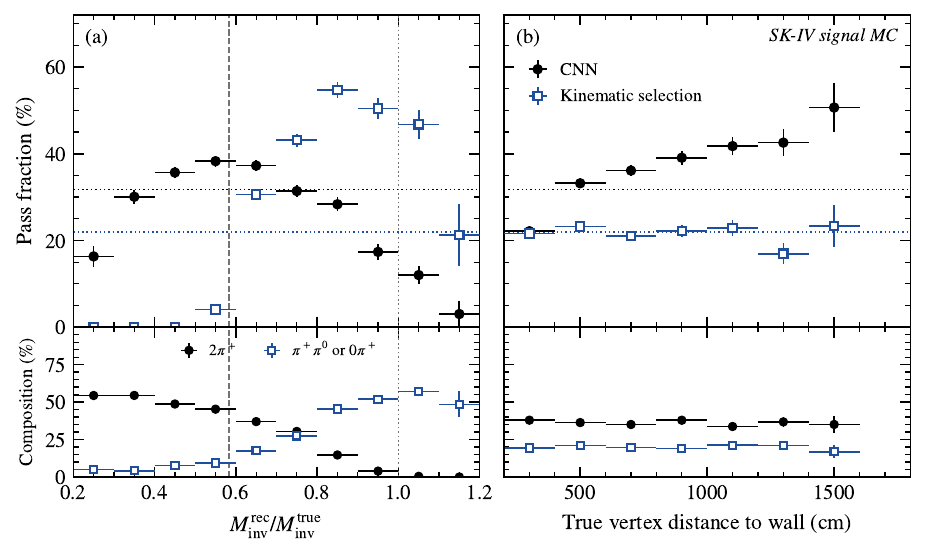}
  \caption{Pass fractions of the CNN (filled circles) and the kinematic selection (open squares) for SK-IV signal-MC events with two to four rings and fewer than three Michel electrons after the precut, as a function of (a) the ratio of reconstructed to true invariant mass and (b) the distance of the true vertex from the wall. Dotted horizontal lines show the overall pass fractions (32\% and 22\%). In (a), the dashed vertical line at $1600/2748=0.58$ marks the ratio below which an event with the median true mass of the $2\pi^+$ class (2,748~MeV/$c^2$) cannot pass the mass window, and the dotted vertical line marks full reconstruction. The lower panels show the fraction of events in the $2\pi^+$ class (filled circles) and in the $\pi^+\pi^0$ or $0\pi^+$ classes (open squares) of Fig.~\ref{fig:ring_eff}(b).}
  \label{fig:pass_curves}
\end{figure*}

Events selected only by the CNN are mostly lost by the kinematic selection because SI of the $\pi^+$ in water cause its energy to be underestimated. A $\pi^+$ above about 500~MeV/$c$ almost always interacts hadronically in water before it stops; part of its energy goes to nucleons below the Cherenkov threshold and its ring is cut short. For matched $\pi^+$ rings, the ratio of the reconstructed momentum to the true $\pi^+$ energy therefore has a median of only 0.32, compared with 1.02 for $e^+$ rings (Fig.~\ref{fig:ring_eff}(a)). For CNN-only events, the true invariant mass of the particles leaving the nucleus (median 2,587~MeV/$c^2$) is close to that of events selected by both methods (2,693~MeV/$c^2$), but the reconstructed $M_{\rm inv}$ is 31\% lower (median 1,322 versus 1,927~MeV/$c^2$), and 84\% of these events fall below the mass window. The CNN uses the hit pattern rather than reconstructed momenta and retains these signal events. Figure~\ref{fig:pass_curves}(a) shows this directly: the kinematic pass fraction rises from zero to 55\% as the ratio of reconstructed to true mass approaches unity, while the CNN pass fraction is highest, about 38\%, at low ratios, where events in the $2\pi^+$ class dominate.

The CNN identifies the target signal topology, in which both $\pi^+$ leave the nucleus and produce rings together with the $e^+$. When FSI absorbs a $\pi^+$, converts it into a $\pi^0$, or produces additional pions, the event no longer has this topology and the CNN output decreases. Most events selected only by the kinematic selection are of this kind: the $\pi^0$ photons produce electromagnetic showers whose energy is well measured, so these events pass the mass window. Compared with events selected by both methods, kinematic-only events more often contain a $\pi^0$ produced in the nucleus (55\% versus 31\%), more than two pions or a $\pi^-$ (25\% versus 6\%), and a knocked-out nucleon (42\% versus 16\%). Figure~\ref{fig:ring_eff}(b) shows the efficiency for each final-state class: for the $2\pi^+$ class (35\% of the sample) the CNN efficiency is 42.5\% compared with 22.0\%, whereas for the $0\pi^+$ class it is 7.8\% compared with 21.4\%. The CNN thus does better on events with the target topology, and the kinematic selection does better on events in which one or more $\pi^+$ have been converted into other particles.

The CNN pass fraction also increases from 22\% to about 50\% with the distance of the true vertex from the wall (Fig.~\ref{fig:pass_curves}(b)), while the kinematic pass fraction stays at about 22\%. The event composition does not depend on the vertex position (lower panel of Fig.~\ref{fig:pass_curves}(b)), and neither does the APFit mass reconstruction: the median ratio of reconstructed to true invariant mass is 0.58--0.62 in every bin of vertex distance, since APFit corrects the charge of each PMT for light attenuation in water and for the PMT angular acceptance, both computed from the reconstructed vertex, before converting it to momentum. The CNN does not use the reconstructed vertex; it takes the uncorrected PMT charge map as input, in which rings from vertices closer to the wall are less completely imaged.

This comparison shows that the two selections use different features of the signal. The CNN identifies the Cherenkov ring pattern of the target topology $e^+\pi^+\pi^+$, while the kinematic selection depends on how completely the final-state energy is reconstructed, which favors events with $\pi^0$ showers. The CNN recovers signal events with the target topology that the kinematic selection loses through SI of the $\pi^+$ in water, giving 1.6 times the efficiency with about one third of the expected background. Because the CNN efficiency relies on events in which both $\pi^+$ leave the nucleus, it decreases when FSI converts a $\pi^+$ into other particles; the CNN efficiency is therefore more sensitive to the pion FSI model, as discussed in Sec.~\ref{sec:systematics}. The two methods are complementary, and combining them may further improve future searches.

\section{Systematic Uncertainty Estimation}
\label{sec:systematics}

We categorize systematic uncertainties into two primary components. The first arises from uncertainties in the physics modeling, which are uncertainties in both the signal and background models.
The second component is closely related to the intrinsic properties of the detector and event reconstruction, which are inherently tied to the CNN model utilized in this analysis. Below is a detailed examination of these uncertainty estimates.

\subsection{Modeling Uncertainties}
Modeling uncertainties are divided into two subcategories: those arising from physical processes and those related to background estimation.

\subsubsection{Physics Modeling Uncertainties}
For the detector model utilized in this study, the largest systematic uncertainty originates from the simulation phase. Across all models, the dominant systematic uncertainty stems from the FSI of pions. These FSI processes significantly affect the Cherenkov ring patterns by modifying the pion kinematics through absorption, charge exchange, and scattering interactions within the nuclear medium. Such modifications can alter the spatial distribution and intensity of the observed Cherenkov rings, potentially impacting our event classification.

To quantify the uncertainty in our analysis, we generated MC simulation samples with 24 parameter perturbations. These perturbations include 1$\sigma$ variations in the $\pi$-nucleon scattering cross-sections---derived from fits to pion scattering experimental data and consistent with the treatment in Ref.~\cite{abe2021}---affecting charge exchange, absorption, inelastic scattering, and hadron production.

\begin{figure}[htbp]
    \centering
    \includegraphics[width=0.9\linewidth]{SK4_Concentration.png}
    \includegraphics[width=0.9\linewidth]{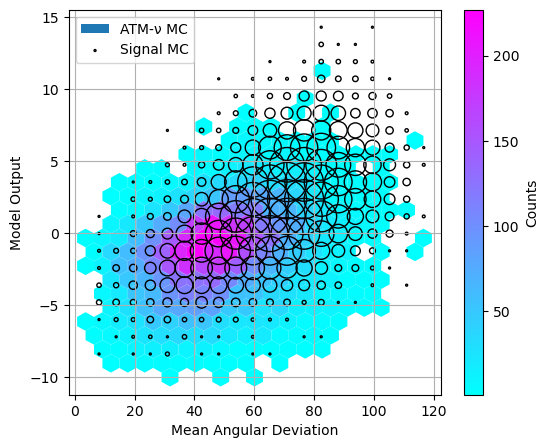}
    \caption{Impact of FSI parameter variations on the mean angular deviation distributions in MC simulations. The top panel shows the distribution using nominal FSI parameters, while the bottom panel shows the distribution with modified FSI parameters including variations in absorption cross sections, scattering processes, and hadron production rates. The mean angular deviation is calculated between the highest-energy particle and other particles emerging from the FSI process for each simulated event.}
    \label{fig:FSIComparison}
\end{figure}

Figure~\ref{fig:FSIComparison} illustrates the impact of these perturbations using the mean angular deviation distribution. 
The top panel shows the distribution with the nominal FSI parameter settings, while the bottom panel highlights the parameter set with the most significant variation among the 24 analyzed: inelastic scattering cross section increased by 50\% at low energies and 28\% at high energies, with accompanying changes including a 50\% decrease in the hadron production cross section, a 36\% increase in pion absorption, and a substantial reduction in charge exchange processes (60\% at low energies and 28\% at high energies). 
These modifications in nuclear interaction probabilities lead to observable shifts in the Cherenkov ring patterns, significantly impacting our signal identification capability.

When evaluating the overall systematic uncertainty, we averaged the changes across all 24 parameters. 
Using the unperturbed CNN model to process the data, we calculated the resulting effects on the signal efficiency and the atmospheric neutrino background rate in the final sample. 
The average efficiency variation observed after applying all analysis selection criteria was adopted as the estimate for this systematic uncertainty. The FSI uncertainty is the dominant systematic in this analysis, amounting to approximately 22--26\% on the signal efficiency and 16--19\% on the atmospheric-neutrino background rate, depending on the SK period. This sensitivity is consistent with the comparison in Sec.~\ref{sec:kinematic_comparison}: the CNN preferentially selects events in which both $\pi^+$ leave the nucleus, and a 10\% relative change in the fraction of such events changes the CNN efficiency by 3.2\% but the kinematic-selection efficiency by only 1.4\%.

In addition, for uncertainties related to Fermi motion and correlated decay, we applied re-weighting techniques to the MC simulations and evaluated the impact on the final signal efficiency and background levels.
Given the incomplete understanding of correlated decay mechanisms, we conservatively assigned a 100\% uncertainty to its detection efficiency. During the simulation phase, we assumed a 10\% occurrence probability for correlated decay and retained this 100\% uncertainty through to the final analysis sample.\cite{takenaka2020}

\subsubsection{Background Modeling Uncertainties}
The uncertainty in background modeling primarily arises from uncertainties in the atmospheric neutrino flux and interaction model. 
Based on previous SK atmospheric neutrino analyses, we constructed a set of weights that describe the behavior of each event under a 1$\sigma$ perturbation in each systematic uncertainty~\cite{abe2018}. 
These weights were applied to MC samples, propagating the uncertainties to the final background prediction. 
Table~\ref{tab:modeling_uncertainties} presents a breakdown of these uncertainties for dominant error sources. 
The uncertainties describe the fractional change in the number of background events under a 1$\sigma$ variation of the source, following the error sources and naming conventions of the SK atmospheric neutrino oscillation analysis~\cite{abe2018}.

\begin{table}[htbp]
    \centering
    \caption{Percentage changes in background event estimates due to systematic uncertainties in the neutrino interaction modeling. Each row corresponds to an independent error source; the deep inelastic scattering (DIS) contributions are itemized separately, following the SK convention~\cite{abe2018}.}
    \label{tab:modeling_uncertainties}
    \setlength{\tabcolsep}{1pt}
    \begin{tabular}{@{}lr@{}}
        \toprule
        \textbf{Source} & \textbf{Uncertainty [\%]} \\ \midrule
        \multicolumn{2}{@{}l}{\textit{Deep inelastic scattering}} \\
        \quad Structure functions & 8.2 \\
        \quad DIS multiplicity & 7.4 \\
        \quad DIS nuclear effects & 6.1 \\
        \quad Low-$W$ DIS/resonance transition & 5.8 \\ \midrule
        Single-meson production (axial mass $M_A$) & 12.3 \\
        Neutrino interaction model & 5.1 \\
        Cross section ratios ($\nu/\bar{\nu}$, $\pi^0/\pi^\pm$) & 4.8 \\ \midrule
        Total systematic uncertainty & 21.2 \\
        \bottomrule
    \end{tabular}
\end{table}

The DIS cross-section modeling provides the largest contribution; its individual error sources---structure functions (8.2\%), hadron multiplicity (7.4\%), nuclear effects (6.1\%), and the low-$W$ DIS/resonance transition (5.8\%)---combine in quadrature to 15.6\%. Additional significant contributions arise from single-meson production via the axial-mass $M_A$ uncertainty (12.3\%), the neutrino interaction model (5.1\%), and the $\nu/\bar{\nu}$ and $\pi^0/\pi^\pm$ cross-section ratios (4.8\%). The combined systematic uncertainty from all sources amounts to 21.2\%.

\subsection{Detector and Reconstruction Uncertainties}
Finally, we considered uncertainties introduced by the detector and event reconstruction. Standard SK analyses treat detector non-uniformity as a source distinct from the absolute and time-dependent energy-scale components. In the SK proton-decay analysis of Ref.~\cite{takenaka2020}, the absolute energy scale is constrained with Michel electrons, reconstructed \(\pi^0\) masses, and stopping cosmic-ray muons; the time variation is combined in quadrature with the absolute component, while zenith-angle-dependent non-uniformity is evaluated separately with Michel electrons. The conventional-fiducial-volume detector non-uniformity uncertainties reported there are 0.6\%, 1.1\%, 0.6\%, and 0.5\% for SK-I through SK-IV, respectively. Ref.~\cite{matsumoto2022} likewise treats detector non-uniformity and energy scale as independent detector/reconstruction sources and propagates them separately to signal efficiency and background. The detector simulation also models the time and position dependence of the water optical properties~\cite{abe2014calib}.

For the present analysis, the detector non-uniformity contribution was evaluated in the established systematic-uncertainty study and is included in the aggregate ``Reconstruction'' entries of Table~\ref{tab:systematic_uncertainties}. These entries, 0.91--1.72\% for signal and 0.82--1.62\% for background, already combine the relevant detector and reconstruction effects for each SK period. We therefore do not add a separate non-uniformity term, which would double count the same contribution.

For the uncertainty assessment of the CNN, we utilized an independent dataset containing control samples, referred to as the ``sideband'' dataset. This  dataset consists of events outside the primary signal region---specifically, topologies not expected to contain the trinucleon decay signal--- that can serve as a control sample to evaluate the CNN’s performance as well as systematic uncertainties in background modeling. 
As shown in Figure~\ref{fig:event_topology_distribution}, the events selected for the sideband analysis focus on specific topologies with two or five Cherenkov rings and one or two decay electrons, highlighted in blue in the figure. 
These sideband regions provide a clean control sample for validating the CNN's performance on event topologies that are distinct from the signal region but share similar detector response characteristics.

\begin{figure}[htbp]
    \centering
    \includegraphics[width=0.9\linewidth]{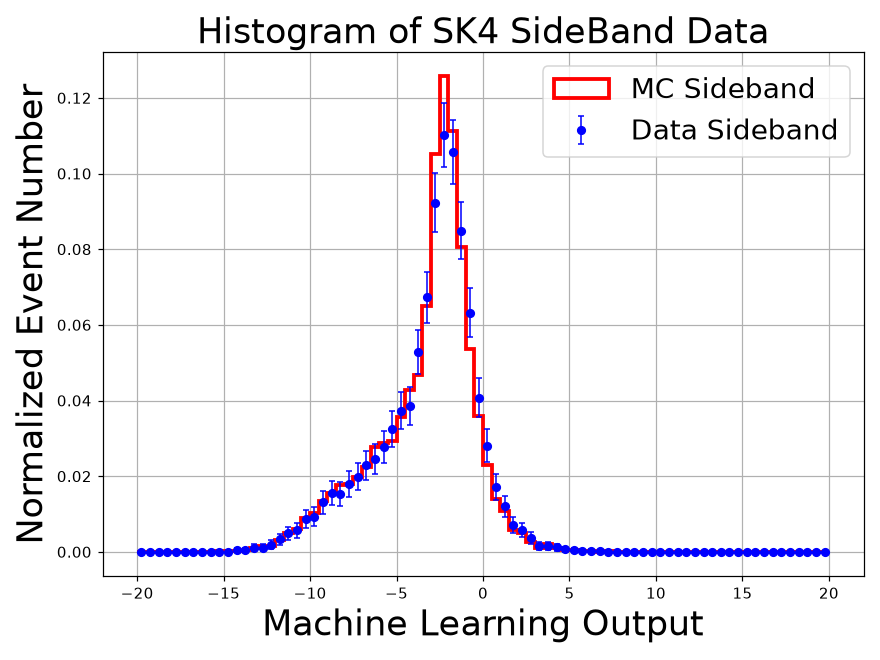}
    \caption{Comparison of CNN output distributions between MC simulations and detector data for SK-IV sideband samples. The red histogram shows atmospheric neutrino MC events, while the blue points represent the corresponding detector data. Both distributions are normalized to unity.}
    \label{fig:SK1to5Comparison}
\end{figure}

By comparing the CNN classification output scores between MC and data in these sideband samples, we quantified the differences in event categorization probabilities. 
Figure~\ref{fig:SK1to5Comparison} illustrates this comparison for the SK-IV period, showing the normalized distribution of CNN output scores for both MC simulated atmospheric neutrino events and actual detector data. The good agreement between simulation and data in these control samples provides confidence in our systematic uncertainty estimation.

For the systematic uncertainty estimation of the CNN output, we adjusted the CNN-score distribution of the MC simulation using a linear transformation of the score axis, \(x \rightarrow ax + b\), where \(x\) represents the CNN output score, to match the detector data distribution in the sideband control sample. The optimal values of \(a\) and \(b\) were determined through a chi-square minimization of the difference between the transformed MC and data distributions. The best-fit transformation $(a,b)$ measures the residual data/MC discrepancy in the shape of the CNN response that is not captured by the nominal simulation. To propagate this into a systematic uncertainty, we apply the same best-fit transformation to the CNN-score distribution of the signal-region MC and recompute the number of signal and background events surviving the CNN cut. The fractional change in these event counts, relative to the nominal (untransformed) MC, is taken as the CNN systematic uncertainty, and is listed in the ``CNN'' row of Table~\ref{tab:systematic_uncertainties}.

\begin{table}[htbp]
    \centering
    \caption{Systematic uncertainties}
    \label{tab:systematic_uncertainties}
    \begin{tabular}{lccccc}
        \toprule
                         & \multicolumn{5}{c}{$^{16}$O(ppp) $\rightarrow$ $^{13}$C $\pi^+ \pi^+ e^+$} \\
        \cmidrule(lr){2-6}
                         & \textbf{SK-I} & \textbf{SK-II} & \textbf{SK-III} & \textbf{SK-IV} & \textbf{SK-V} \\
        \cmidrule(lr){2-6}
        \multicolumn{6}{l}{\textbf{Signal (\%)}} \\
        \midrule
        Simulation       & 25.92 & 25.61 & 25.41 & 23.95 & 24.95 \\
        Reconstruction   & 1.2   & 1.72  & 1.1   & 0.91  & 1.1   \\
        CNN              & 6.7   & 8.8   & 8.4   & 6.7   & 8.2   \\
        \textbf{Total}   & \textbf{26.80} & \textbf{27.13} & \textbf{26.79} & \textbf{24.89} & \textbf{26.29} \\
        \midrule
        \multicolumn{6}{l}{\textbf{Background (\%)}} \\
        \midrule
        Simulation       & 26.09 & 28.24 & 24.86 & 27.45 & 25.37 \\
        Reconstruction   & 1.22  & 1.62  & 0.82  & 0.91  & 1.1   \\
        CNN              & 6.7   & 8.8   & 8.4   & 6.7   & 8.2   \\
        \textbf{Total}   & \textbf{26.96} & \textbf{29.62} & \textbf{26.25} & \textbf{28.27} & \textbf{26.68} \\
        \bottomrule
    \end{tabular}
\end{table}

The calculated systematic uncertainties for signal and background events across all SK periods are summarized in Table~\ref{tab:systematic_uncertainties}, showing contributions from simulation, reconstruction, and CNN analysis. The uncertainties remain consistent across different SK periods, with total uncertainties ranging from 24.89\% to 27.13\% for signal events and 26.25\% to 29.62\% for background events.

\section{Results}
\label{sec:results}
In this section, we analyze data collected from SK-I to SK-V, with a cumulative exposure of $0.401$ megaton-years. Applying the selection criteria established in Section~\ref{sec:ml_analysis}, we processed the full dataset through our analysis chain. Figure~\ref{fig:CNNOutputLog} shows representative results from the SK4 period, demonstrating the signal-background separation achieved by our selection. The analysis results for other SK periods showed similar characteristics and can be found in the appendix.

\begin{figure}[htbp]
  \centering
  \includegraphics[width=\columnwidth]{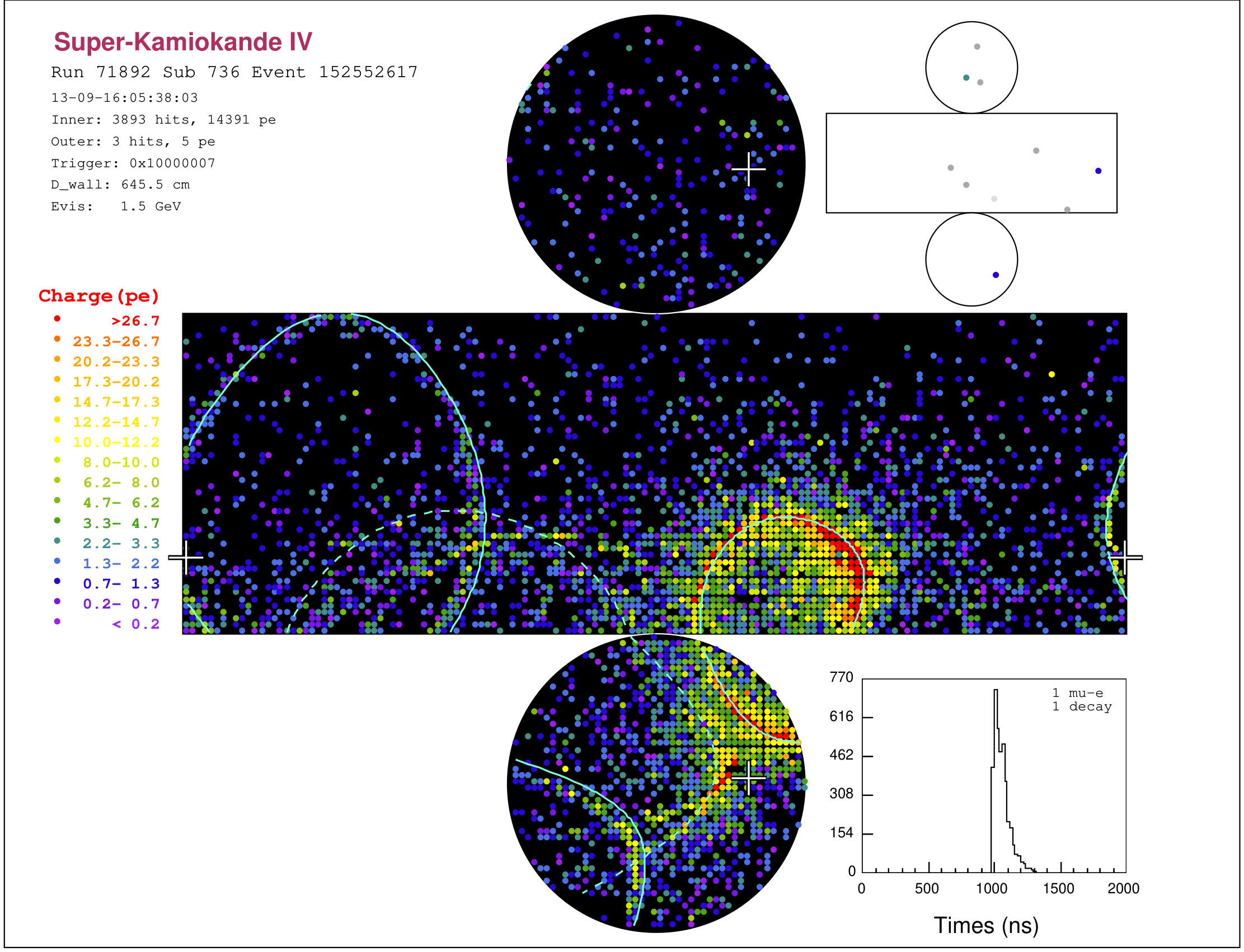}
  \caption{Event display of the identified event in the SK detector, showing the event location at the detector center. The figure also includes the reconstructed result using traditional methods, illustrating the particle tracks and their respective Cherenkov rings.}
  \label{fig:EventDisplay}
\end{figure}

After applying both the pre-selection and CNN-based selections, only one event remained in the signal region.
Figure~\ref{fig:EventDisplay} shows the event display of the final candidate. 
This event was observed during SK-IV operations, with a CNN prediction value of 6.20, marginally exceeding the period-specific cut threshold of 6.12.
The observation of one event is consistent with an expected background of approximately $1.33$ events, as summarized in Table~\ref{tab:eff_bkg_data_transposed}. 
The Poisson probability of observing one event, given this background expectation is 35.2\%.

\begin{figure*}[!t]
  \centering
  \includegraphics[width=0.9\textwidth]{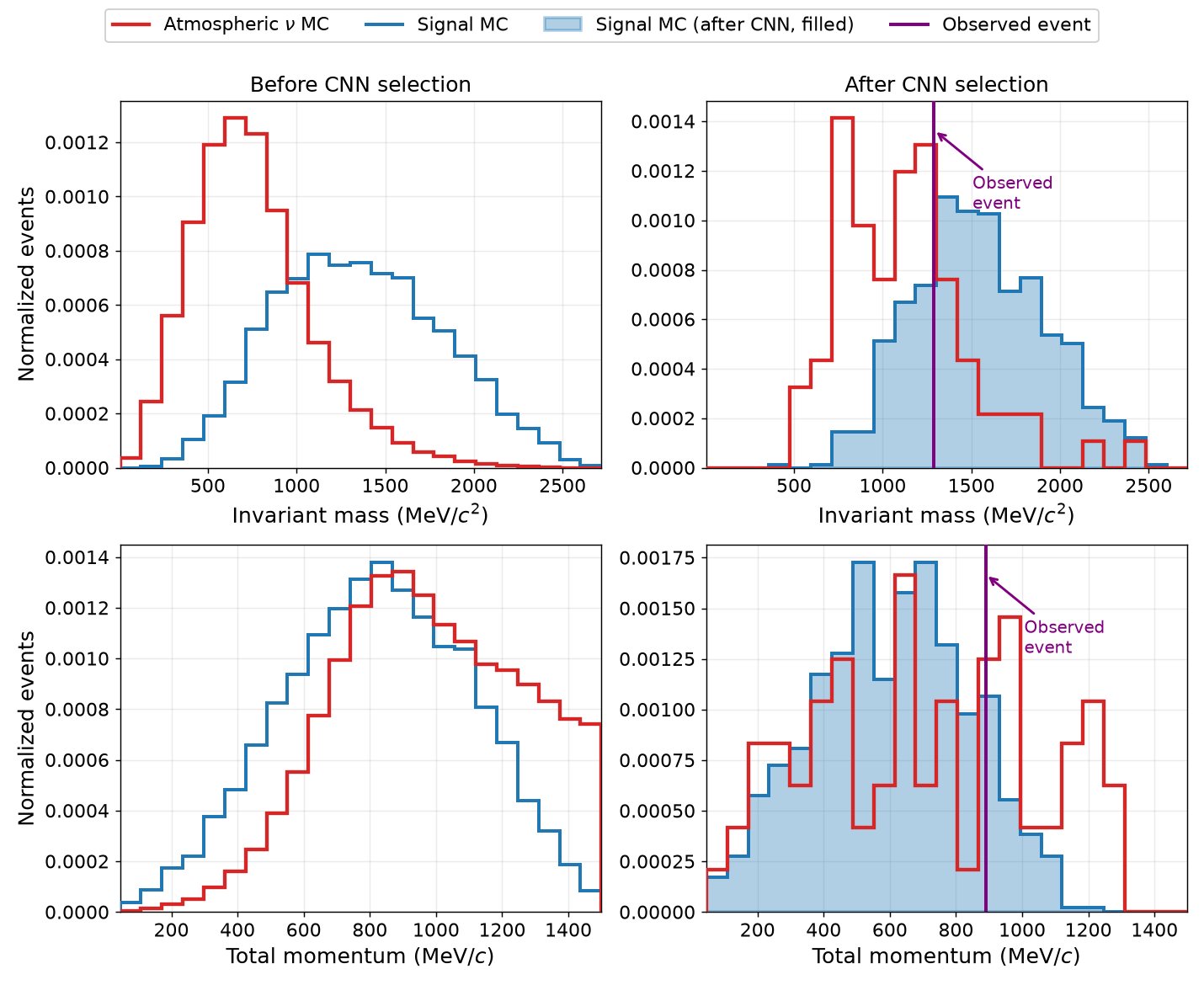}
  \caption{Distributions of the reconstructed invariant mass (top row) and total momentum (bottom row) from the traditional reconstruction, shown separately before (left column) and after (right column) the CNN selection. In each panel the atmospheric-neutrino background MC is shown in red and the trinucleon-decay signal MC in blue (filled after the CNN selection). Before the CNN selection the two distributions overlap strongly, whereas after the selection the surviving signal MC is shifted to higher invariant mass and lower total momentum, clearly separated from the surviving background. All histograms are area-normalized to unity so that the change in \emph{shape} produced by the CNN selection is emphasized rather than the (much larger) reduction in absolute background rate. The observed candidate event is indicated by the vertical purple line in the right-hand panels.}
  \label{fig:combined_distributions}
\end{figure*}

To cross-validate these results, we employed traditional reconstruction methods, with the resulting invariant mass and total momentum distributions shown in Figure~\ref{fig:combined_distributions}. 
This analysis follows methodologies established in previous SK studies~\cite{Shiozawa1999, ashie2005, mine2024}, where each Cherenkov ring is reconstructed to determine particle type and momentum. However, the traditional reconstruction approach reveals several notable limitations. 
We obtained a reconstructed invariant mass of 1.2 GeV/c\textsuperscript{2} and a total momentum of 888 MeV/c. 
While these values fall within the broader distribution of signal MC events, they deviate notably from the expected central values. 
The reconstructed invariant mass is somewhat lower than the expected peak around 1.5 GeV/c\textsuperscript{2}, and the total momentum, though within the allowed range, differs from the central values predicted by our MC simulations. Furthermore, all three reconstructed particles exhibited electron-like signatures, which contrasts with the expected signature of two \(\pi^+\) and one \(e^+\), where we would anticipate observing two muon-like rings and one electron-like ring.

\begin{table}[htbp]
  \centering
  \caption{Comparison of trinucleon decay lifetime limits at 90\% confidence level between this work and previous experiments~\cite{gerda2023, alvis2019}.}
  \label{tab:lifetime_limits}
  \begin{tabular}{lc}
    \toprule
    \textbf{Experiment} & \textbf{Lifetime Limit (90\% C.L.)} \\
    \midrule
    Super-Kamiokande     & $\tau > 4.2 \times 10^{32}$ years \\
    GERDA                & $\tau > 4.7 \times 10^{25}$ years \\
    Majorana Demonstrator& $\tau > 1.6 \times 10^{26}$ years \\
    \bottomrule
  \end{tabular}
\end{table}

We employ a Bayesian method incorporating systematic uncertainties to set an upper limit for the trinucleon decay rate. 
The formula for the posterior probability distribution is given by
\begin{equation}
\begin{aligned}
  P(\Gamma | n_i) = \iiint
  &\frac{e^{- (\Gamma \lambda_i \epsilon_i + b_i)} (\Gamma \lambda_i \epsilon_i + b_i)^{n_i}}{n_i!} \\
  &\times P(\Gamma) P(\lambda_i) P(\epsilon_i) P(b_i) \, d\lambda_i \, d\epsilon_i \, db_i,
\end{aligned}
\end{equation}
\noindent where the index $i$ runs over the five SK run periods (SK-I--SK-V); $\Gamma$ is the true trinucleon decay rate, and $n_i$, $\lambda_i$, $\epsilon_i$, and $b_i$ are, respectively, the number of candidate events, the true exposure, the true signal efficiency, and the true background rate for the $i$th period. The decay-rate prior $P(\Gamma)$ is uniform for $\Gamma \geq 0$ and zero otherwise, while $P(\lambda_i)$, $P(\epsilon_i)$, and $P(b_i)$ are Gaussian priors whose widths are set by the corresponding systematic uncertainties; the exposure uncertainty is taken to be negligible.

The five run periods are combined by taking the product of their per-period posterior distributions. The 90\% confidence-level upper limit on the decay rate, $\Gamma_{\text{limit}}$, is defined by
\begin{equation}
  \mathrm{C.L.} = \frac{\displaystyle\int_{0}^{\Gamma_{\text{limit}}} \prod_{i=1}^{5} P(\Gamma | n_i)\, d\Gamma}{\displaystyle\int_{0}^{\infty} \prod_{i=1}^{5} P(\Gamma | n_i)\, d\Gamma}.
\end{equation}

At the 90\% confidence level, we establish a lower limit on the lifetime of $\tau > 4.2 \times 10^{32}$ years for this decay mode. Table~\ref{tab:lifetime_limits} presents our results alongside previous experimental limits set by the GERDA and Majorana Demonstrator collaborations~\cite{gerda2023, alvis2019}.

\section{Conclusion}
\label{sec:conclusion}

In conclusion, utilizing data from SK-I to SK-V, we conducted a search for trinucleon decay using deep learning methods for the first time in the SK experiment. 

We observed only one event, which is consistent with an expected background of $1.33$ events, showing no statistically significant excess. Consequently, we set a lower limit on the trinucleon decay lifetime at $\tau > 4.2 \times 10^{32}$ years. This result achieves more than a six-order-of-magnitude improvement over previous limits and reaches the theoretical model prediction, which estimates the partial lifetime of this $\Delta B = 3$ mode to be on the order of $10^{33}$ years within the anomaly-free $Z_6$ baryon-parity framework~\cite{babu2003}.

\section*{Data Availability}
The numerical data needed to reproduce the figures in this paper are available from the corresponding author upon reasonable request. These data are limited to the numerical values used in the figures and do not include the full Monte Carlo event samples. The classification analysis uses the publicly available PyTorch framework~\cite{pytorch} and the MobileNetV3 pretrained weights~\cite{huggingface_mobilenetv3}.

\section{Acknowledgments}

We gratefully acknowledge the cooperation of the Kamioka Mining and Smelting Company.
The Super-Kamiokande experiment has been built and operated from funding by the 
Japanese Ministry of Education, Culture, Sports, Science and Technology; the U.S.
Department of Energy; and the U.S. National Science Foundation. Some of us have been 
supported by funds from the National Research Foundation of Korea (NRF-2009-0083526,
NRF-2022R1A5A1030700, NRF-2022R1A3B1078756) funded by the Ministry of Science, 
Information and Communication Technology (ICT); the Institute for 
Basic Science (IBS-R016-Y2); and the Ministry of Education (2018R1D1A1B07049158,
2021R1I1A1A01042256, 2021R1I1A1A01059559, RS-2024-00442775);
the Japan Society for the Promotion of Science; the National
Natural Science Foundation of China under Grants No.12375100; the Spanish Ministry of Science, 
Universities and Innovation (grant PID2021-124050NB-C31); the Natural Sciences and 
Engineering Research Council (NSERC) of Canada; the Scinet and Westgrid consortia of
Compute Canada; 
the National Science Centre (UMO-2018/30/E/ST2/00441 and UMO-2022/46/E/ST2/00336) 
and the Ministry of  Science and Higher Education (2023/WK/04), Poland;
the Science and Technology Facilities Council (STFC) and
Grid for Particle Physics (GridPP), UK; the European Union's 
Horizon 2020 Research and Innovation Programme under the Marie Sklodowska-Curie grant
agreement no.754496; H2020-MSCA-RISE-2018 JENNIFER2 grant agreement no.822070, H2020-MSCA-RISE-2019 SK2HK grant agreement no. 872549; 
and European Union's Next Generation EU/PRTR  grant CA3/RSUE2021-00559; 
the National Institute for Nuclear Physics (INFN), Italy.

OpenAI Codex (model \texttt{gpt-6-astra}) was used to assist with the manuscript's English and for arithmetic consistency checks of the exposure-normalized background counts. The authors retain responsibility for the scientific content and final manuscript.

\clearpage
\onecolumngrid
\section*{Appendix: Figures for SK1–SK5}  %

\begin{center}
  \includegraphics[width=0.46\textwidth]{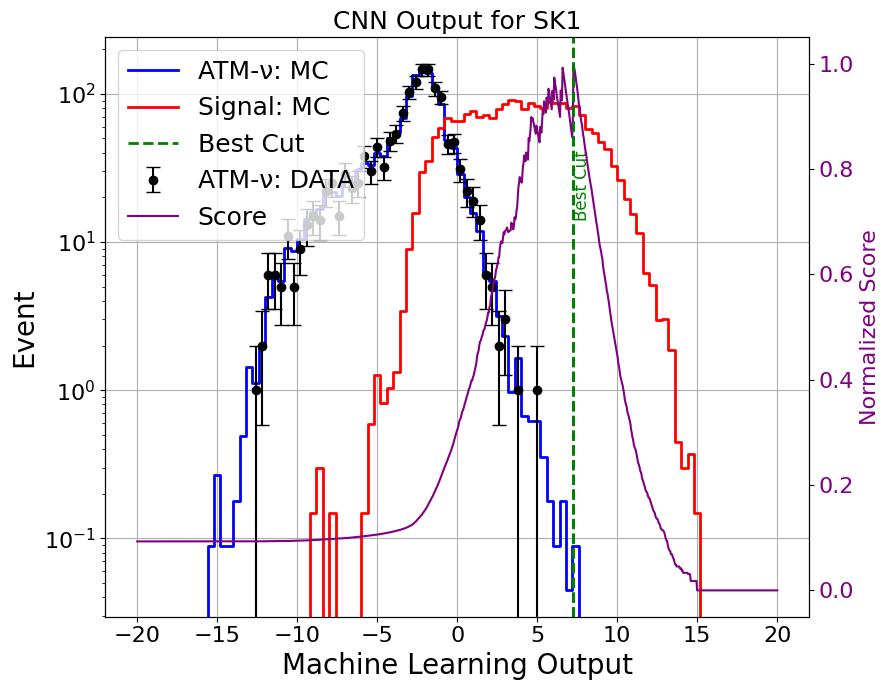}\hfill
  \includegraphics[width=0.46\textwidth]{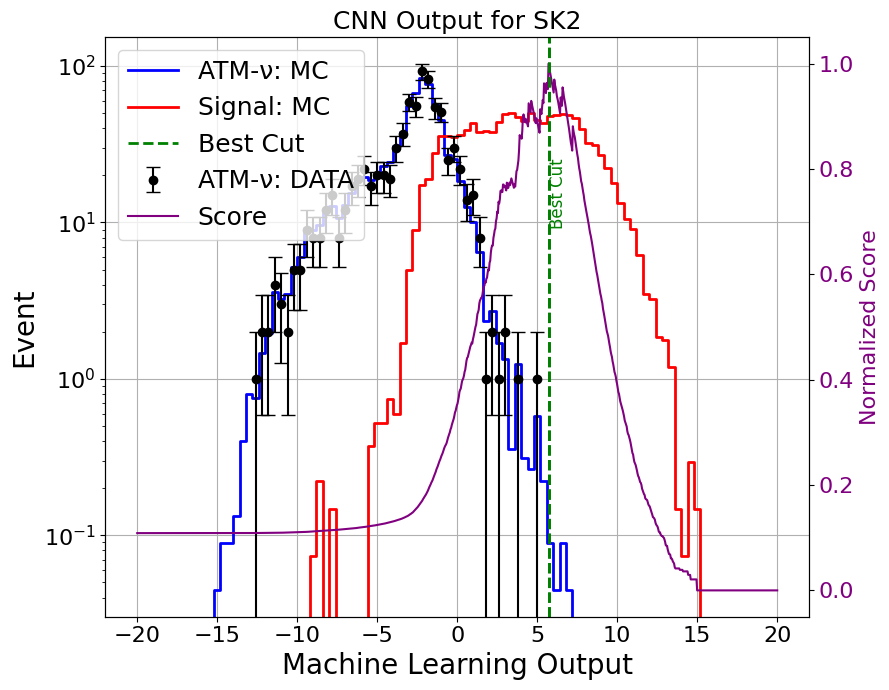}\\[4pt]
  \includegraphics[width=0.46\textwidth]{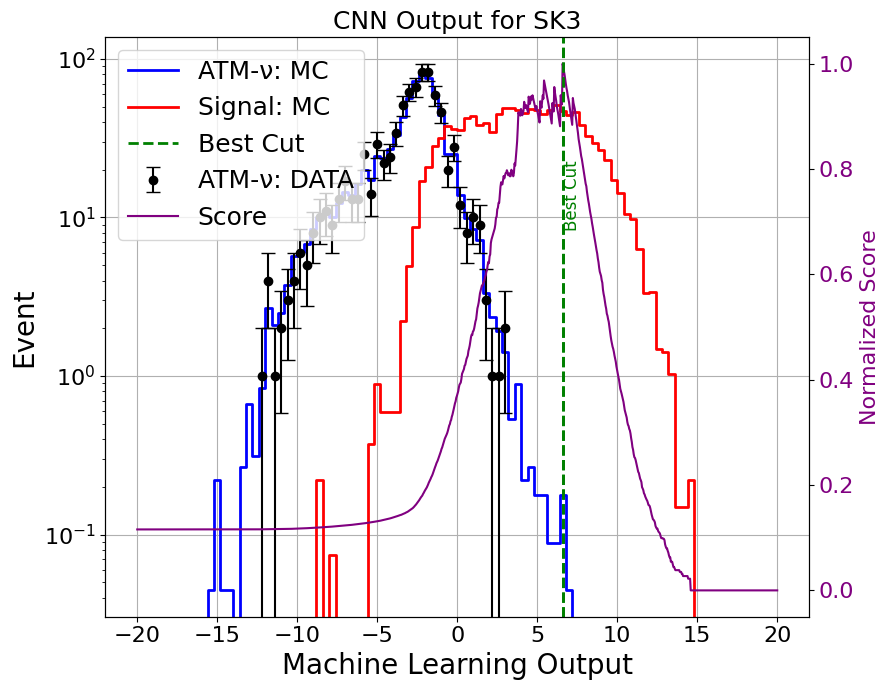}\hfill
  \includegraphics[width=0.46\textwidth]{cnn_output_sk4_log.png}\\[4pt]
  \includegraphics[width=0.46\textwidth]{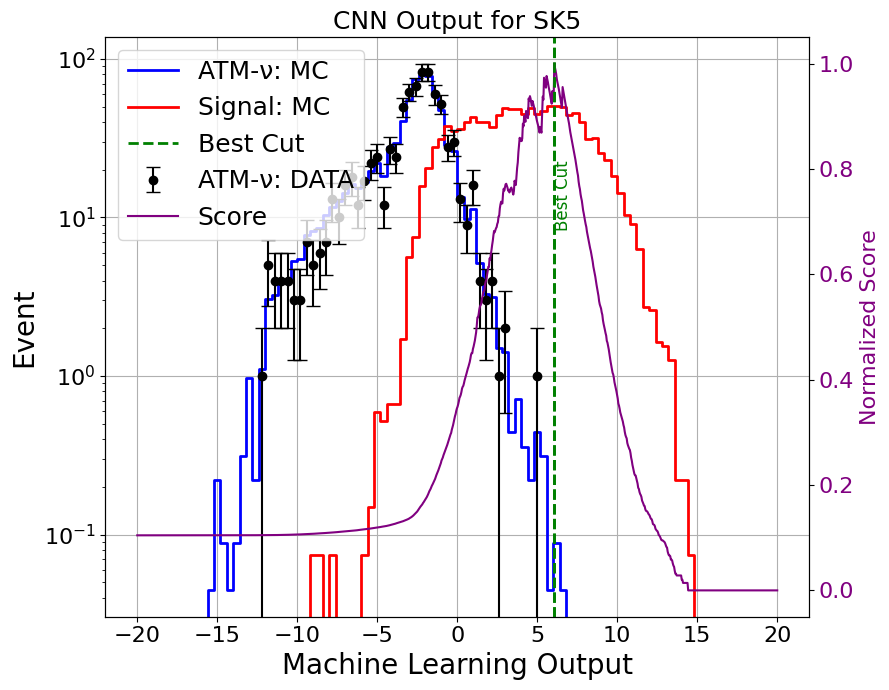}
  \captionof{figure}{Machine learning output distributions (logarithmic scale) for the SK-I through SK-V periods. Blue: atmospheric-neutrino MC; red: trinucleon-decay signal MC; black points: data; green dashed line: optimal cut; purple curve: normalized score.}
  \label{fig:appendix_log}
\end{center}
\clearpage

\begin{center}
  \includegraphics[width=0.32\textwidth]{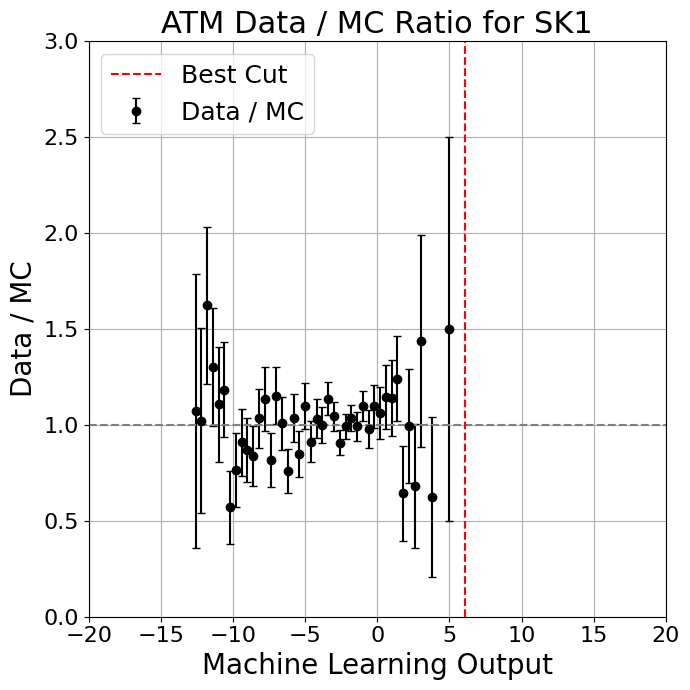}\hfill
  \includegraphics[width=0.32\textwidth]{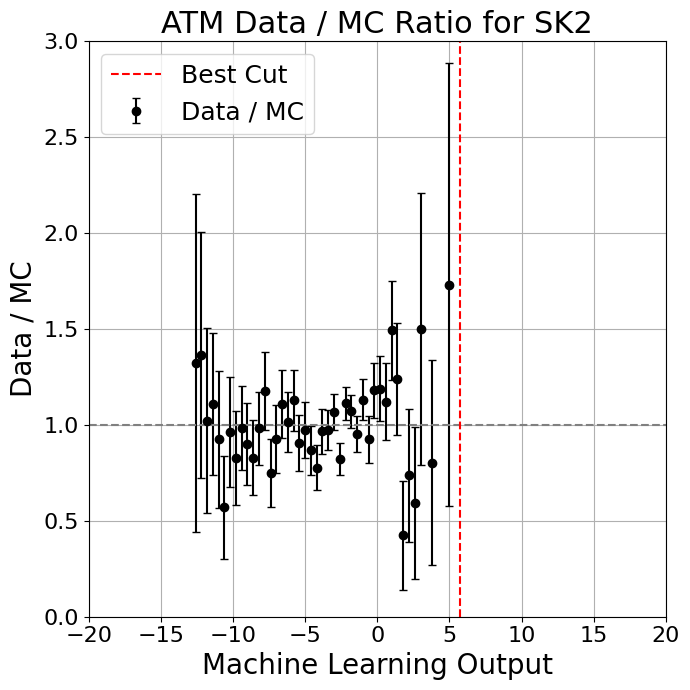}\hfill
  \includegraphics[width=0.32\textwidth]{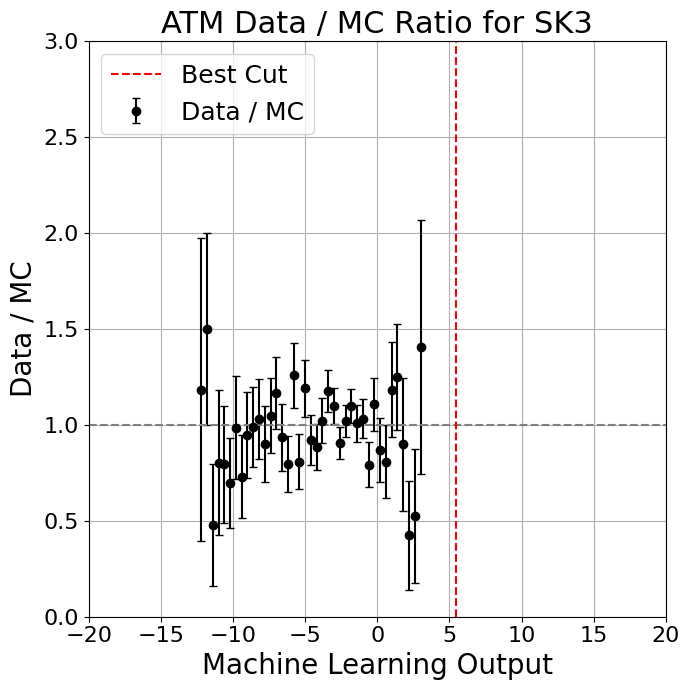}\\[4pt]
  \includegraphics[width=0.32\textwidth]{cnn_output_sk4_ratio1.png}\hspace{0.02\textwidth}
  \includegraphics[width=0.32\textwidth]{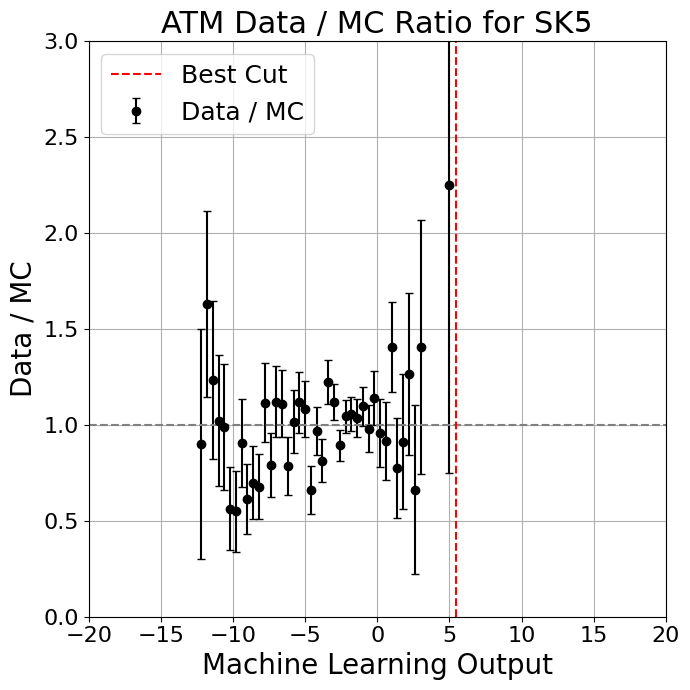}
  \captionof{figure}{Atmospheric-neutrino data/MC ratio of the machine learning output for the SK-I through SK-V periods. The red dashed line indicates the optimal cut.}
  \label{fig:appendix_ratio}
\end{center}
\twocolumngrid

\end{document}